\documentclass[journal]{IEEEtran}
\IEEEoverridecommandlockouts
\usepackage{cite}
\usepackage{amsmath,amssymb,amsfonts}
\usepackage{algorithmic}
\usepackage{graphicx}
\usepackage{textcomp}
\usepackage{xcolor}
\def\BibTe\checkmark{{\rm B\kern-.05em{\sc i\kern-.025em b}\kern-.08em
    T\kern-.1667em\lower.7ex\hbox{E}\kern-.125em\checkmark}}
\usepackage{array, colortbl}
\usepackage{ulem, comment}
\usepackage{longtable}
\usepackage{adjustbox}
\usepackage{slashbox}
\usepackage{csquotes}
\renewcommand{\mkbegdispquote}[2]{\itshape}
\usepackage{soul,color}
\usepackage{tabularray}
\usepackage[justification=centering]{subfig}
\usepackage[many]{tcolorbox}
\usepackage{balance}
\usepackage{mathtools}
\usepackage{multirow}
\usepackage{subfig}
\DeclareMathOperator*{\argmax}{arg\,max}

\newcommand{\xMapsto}[2][]{\ext@arrow 0599{\Mapstofill@}{#1}{#2}}
\def\Mapstofill@{\arrowfill@{\Mapstochar\Relbar}\Relbar\Rightarrow}
\newcommand{\PHI}[0]{\operatorname*{\phi}}

\usepackage[author={Anonymous}]{pdfcomment}

\newcommand{\etal}{\textit{et al.}}
\newtcolorbox{boxblock}[2][]{
top=0.15in,left=4pt,right=4pt,bottom=4pt,
fonttitle=\bfseries,
colbacktitle=gray,
colback=gray!5,
colframe=gray!40!black,
enhanced,
attach boxed title to top left={xshift=1.5em,yshift=-\tcboxedtitleheight/2},
boxed title style={size=small},
drop shadow={black!50!white},
title=#2,#1}

\begin{document}

\title{Reliable Malware Analysis and Detection using Topology Data Analysis
}

\author{Lionel~Nganyewou~Tidjon and Foutse~Khomh\IEEEmembership{, Senior Member,~IEEE}
\IEEEcompsocitemizethanks{\IEEEcompsocthanksitem The authors are with Polytechnique Montréal, Montréal, QC H3C 3A7, Canada (e-mail: lionel.tidjon@polymtl.ca, foutse.khomh@polymtl.ca).
}}


\maketitle

\begin{abstract}
Increasingly, malwares are becoming complex and they are spreading on networks targeting different infrastructures and personal-end devices to collect, modify, and destroy victim information. Malware behaviors are polymorphic, metamorphic, persistent, able to hide to bypass detectors and to adapt to new environments, and even leverage machine learning techniques to better damage targets. Thus, it makes them difficult to analyze and detect with traditional endpoint detection and response, intrusion detection and prevention systems. To defend against malwares, recent work have proposed different techniques based on signatures and machine learning. In this paper, we propose to use an algebraic topological approach called topological-based data analysis (TDA) to efficiently analyze and detect complex malware patterns. Next, we compare the different TDA techniques (i.e., persistence homology, tomato, TDA Mapper) and existing techniques (i.e., PCA, UMAP, t-SNE) using different classifiers including random forest, decision tree, xgboost, and lightgbm. We also propose some recommendations to deploy the best-identified models for malware detection at scale. 
Results show that TDA Mapper (combined with PCA) is better for clustering and for identifying hidden relationships between malware clusters compared to PCA. Persistent diagrams are better to identify overlapping malware clusters with low execution time compared to UMAP and t-SNE. For malware detection, malware analysts can use Random Forest and Decision Tree with t-SNE and Persistent Diagram to achieve better performance and robustness on noised data. 

\end{abstract}


\section{Introduction}
\label{intro}


\IEEEPARstart{N}{owadays}, attackers leverage complex tactics, techniques, and procedures (TTPs) to spread malwares and bypass intrusion detection/prevention systems~\cite{8735821}. Malware goals include gaining access to the target system, hiding on the system to spy user activities (e.g., phone calls, SMS), gathering and exfiltrating personal information (e.g., bank accounts, card information, documents), disrupting the target network making it unavailable, and destroying user information. 

\begin{figure}[]
\centering
\includegraphics[scale=0.4]{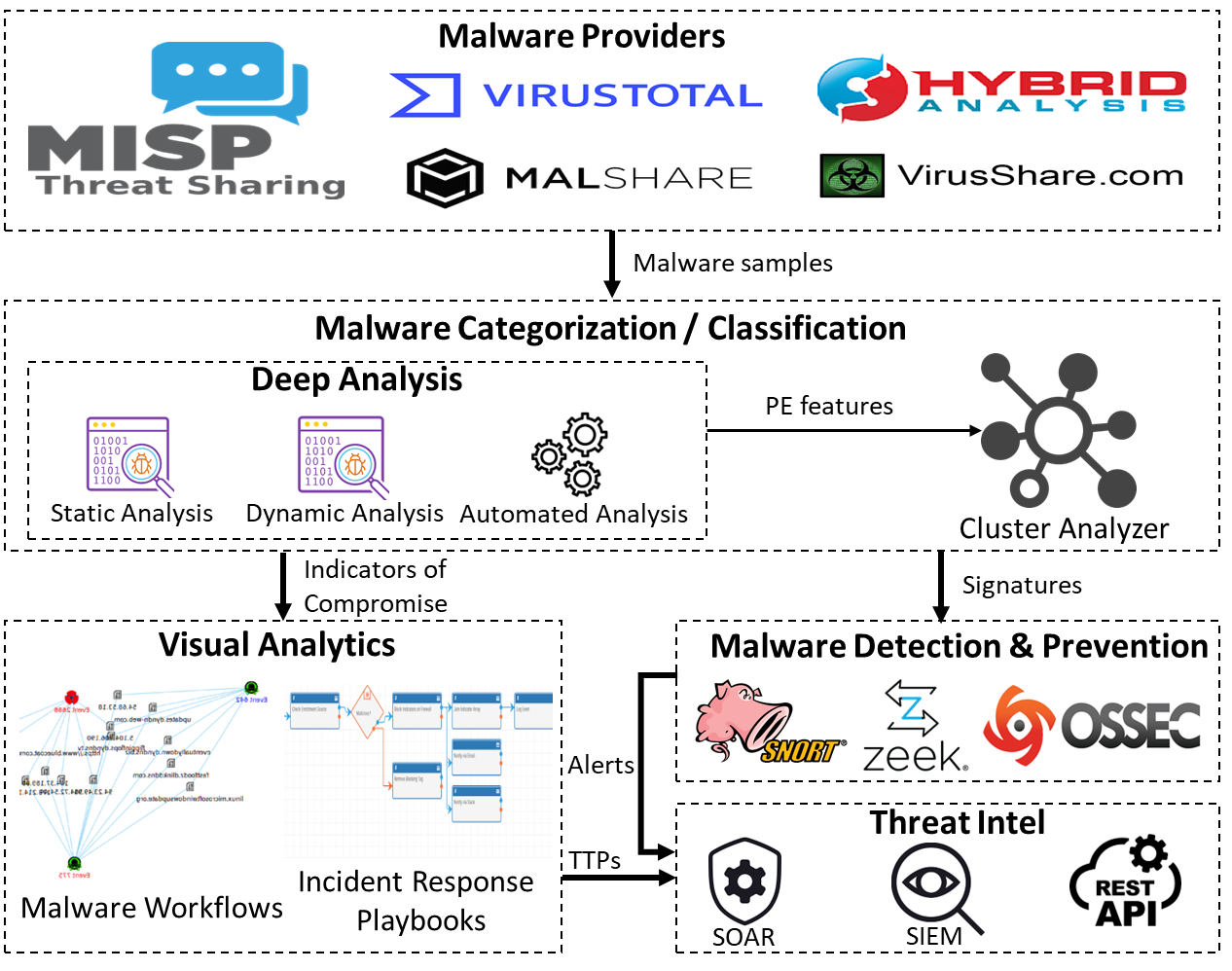}
\caption{Malware Analysis Architecture}
\label{fig00}
\vspace{-0.5em}
\end{figure}

Malware analysis helps security analysts to investigate and understand the intentions behind a malicious code sample; thus allowing to take proper courses of action and react to zero-day threats. In Security Operations Centers, malware samples are often ingested from various external providers such as Virus Total, Hybrid Analysis, and MISP Threat Sharing (see Fig.~\ref{fig00}). Then, malware samples are analyzed using Portable Executable (PE) features embedded in malware binaries such as operational codes (opcodes), APIs/system calls, CPU registers, network calls, string signatures, and file characteristics (e.g., hash). Usually, these features can be identified using three approaches: static analysis, dynamic analysis, and hybrid analysis (combining static and dynamic analysis). Static analysis consists in scrutinizing malicious code instructions and memory in the malware binary without executing it. In an emulated/virtual environment, dynamic analysis identifies malicious patterns while the malware binary is being executed. The analysis process is often automated, semi-automated, or manual. Next, a cluster analyzer checks relationships between malware clusters. Signatures extracted from malware clusters are used to feed intrusion detection and prevention systems (e.g., snort, zeek). In addition, Indicators of Compromise (IoCs) help to visually understand malware TTPs and to respond to each attack step using Incident Response playbooks. TTPs and alerts from malware detectors are consumed by threat intelligence platforms (e.g., SOAR) and APIs for security management, automation, and response. 

To deal with a large number of malwares, analysis and detection tools must satisfy the following requirements: be accurate at identifying malware categories, provide automation to reduce the analysis effort, have a low execution time, a low memory usage, provide reproducible analysis to allow different security teams~\cite{daoudi2021lessons} to repeat a malware analysis, and be robust to noised data like network traffic or logs (e.g., in the case of a network/host malware detector). Currently, cyberanalysts manually write low-level rules based on identified IoCs for pattern matching (e.g., yara, sigma) and correlation. This approach precisely identifies known malware samples with a low false positive rate but it becomes ineffective against polymorphic/metamorphic malwares with complex pattern mutations and zero-day malwares. It also requires a high amount of effort for malware analysts to rewrite new rules per new malware sample. Machine learning has been used to overcome these limitations by enabling automatic identification of anomalous malware patterns but it often generates many false positives on real-world data~\cite{8735821}. Machine learning also suffers from reproducibility and robustness issues~\cite{daoudi2021lessons, tidjon2022threat}. 

In this paper, we propose to use topological data analysis (TDA)~\cite{edelsbrunner2000topological,zomorodian2005computing,singh2007topological,chazal2013persistence, carriere2021optimizing} for malware analysis and detection to efficiently assist malware analysts in their daily tasks. This choice is motivated by TDA particularities such as easier data exploration using topological graphs and persistence~\cite{edelsbrunner2000topological}, guarantees of the robustness under small perturbations via persistence~\cite{cohen2007stability}, 
reproducibility~\cite{hajij2021persistent}, efficiency for data classification (e.g., infectious diseases~\cite{chen2021topological, bleher2021topology}, oncology~\cite{aukerman2020persistent, nicolau2011topology, oyama2019hepatic}, renewable energy~\cite{bush2021topological}), and automation. We experiment with different TDA techniques such as persistence homology~\cite{edelsbrunner2000topological, zomorodian2005computing, ghrist2008barcodes, chazal2009gromov, bubenik2015statistical, adams2017persistence}, TDA Mapper~\cite{singh2007topological}, and tomato~\cite{chazal2013persistence} with the aim to efficiently analyze and detect complex malwares under/without perturbations while being able to get the same results for different runs with a minimal overhead. The experimentation is based on two datasets from the Canadian Institute for Cybersecurity (CIC): CIC-MalMem-2022 and CCCS-CIC-AndMal-2020. We also propose some recommendations to deploy the best-identified models for malware detection at scale. Results show that (1) TDA Mapper (combined with PCA) is better for clustering and to identify hidden relationships between malware clusters compared to PCA, (2) persistent diagrams are better to identify overlapping malware clusters with low execution time compared to UMAP and t-SNE, (3) Random Forest and Decision Tree with t-SNE and Persistent Diagram are more performant and robust on data with too much noise, and (4) XGBoost and LightGBM can only be used with t-SNE to achieve better performance and robustness. Thus, they can be effectively used by malware analysts to improve their regular malware analysis and detection practices.

The rest of this paper is structured as follows.
In Section~\ref{related-work}, we review the related literature. Section~\ref{back} presents concepts and formal definitions of topological data analysis (TDA).
The study methodology to evaluate TDA techniques for malware analysis and detection is shown in Section~\ref{methodology}. In Section~\ref{evaluation}, we present results while answering to the defined research questions. Section~\ref{discussions} discusses the results obtained.
Section~\ref{recommendations} proposes some recommendations to deploy the best-identified models for malware detection at scale.
In Section~\ref{threat2valid}, we discuss threats that could affect the validity of the reported results.
Section~\ref{conclusion} concludes the paper and outlines avenues for future work.

\section{Related work}
\label{related-work}

In this section, we review recent work on the application of Topological Data Analysis (TDA) to cybersecurity. 

Akcora \etal{}~\cite{10.5555/3491440.3492052} used TDA Mapper to recover hidden data patterns for Bitcoin malware detection that are inaccessible using conventional data analytic techniques like t-SNE. They analyzed different ransomwares (e.g., Cerber, CryptoLocker) and successfully identified unique suspicious addresses and true ransomware addresses. The authors only used TDA Mapper and did not consider other TDA techniques like persistent diagrams and Tomato. 

R.J. Gutierrez \etal{}~\cite{gutierrez2018cyber} proposed a tabulated vector approach to process big firewall log files and identify anomalies within the flagged firewall log event data from regional data nodes like Microsoft, Verizon, and AT\&T Network Operations Center. For graph analysis, they analyzed IP network graphs using histogram matrix and TDA mapper to quickly identify the IP dynamics within the firewall log events. In this paper, we do not use histogram matrix but we rather explore all TDA techniques such as TDA mapper, Tomato, and persistence homology for malware detection.

In~\cite{davies2022topological}, the author used persistent diagrams to identify anomalies in host-based logs captured from the Logging Made Easy project and he observed that the structure captured by persistent diagrams contained discriminative information on whether the logs are anomalous. In this work, we do not only use persistent diagrams but we also use other TDA techniques (TDA mapper, Tomato) for malware analysis and detection on different environments (mobile, desktop, network). 

N. Masaki \etal{}~\cite{narita2021empirical} used TDA Mapper to recognize malicious packets on the darknet acquired by an Internet threat monitoring system from JPCERT Coordination Center. P. Bruillard \etal{}~\cite{bruillard2016anomaly} applied persistence homology to analyze Netflow data~\cite{shiravi2012toward} for anomalous traffic detection. The Netflow data consisted of four attack scenarios: network penetration and backdoor setup, DoS, IRC botnet, and brute-force SSH. This paper extends the work done in~\cite{narita2021empirical, bruillard2016anomaly, shiravi2012toward} by analyzing the application of all TDA techniques for malware detection on multiple datasets (balanced, imbalanced). The paper also proposes some recommendations for malware analysis and detection to efficiently improve their daily malware analysis tasks and malware detectors. To the best of our knowledge, there is no such work in the literature. In the following, we define TDA concepts and techniques.   

\section{Background}
\label{back}
This section presents formal definitions of the following TDA techniques : TDA Mapper, persistence homology, and tomato. 

\subsection{Definitions}
A detailed introduction about the theoretical foundations of TDAs can be found in~\cite{hatcher2005algebraic, wasserman2016topological, chazal2021introduction}.

\textbf{Metric spaces.} Let $(X, d_{X})$ be a metric space where $X$ is a set and $d_{X}: X \times X \rightarrow \mathbb{R}^{+}$ is a distance function such that for any $x,y,z \in X$,
\begin{itemize}
    \item[(1)] $d_{X}(x, y) = 0$ if and only if $x=y$,
    \item[(2)] $d_{X}(x, y) = d_{X}(y, x)$,
    \item[(3)] $d_{X}(x, z) \leq d_{X}(x, y) + d_{X}(y, z)$
\end{itemize}

In the real metric space $(\mathbb{R}^{n}, ||.||)$, we consider $X = \{ x_0,x_1,...,x_k\}$ be a finite set of points such that $X \subseteq \mathbb{R}^{n}$ and for any $x_i, x_j\in X, d_{X}(x_i, x_j) = || x_i - x_j ||$. The $k$-simplex $\sigma = [x_0,...,x_k]$ spanned by $X$ is the convex hull of $X$, the points $(x_i)_{i \in \{0,..,k\}}$ are the vertices of $\sigma$ and the simplices spanned by the subsets of $X$ are the faces of $\sigma$.

\textit{Example 1.} Let $X=\{1,2,3\}$ be a set that form a triangle. $\sigma=[1,2,3]$ is the convex hull of the triangle and $\{1,2\}$, $\{2,3\}$, $\{1,3\}$ are some instances of the faces of $\sigma$.

\textbf{Simplicial complexes.} A simplicial complex with the vertex set $X$ is defined by a collection $K$ in $\mathbb{R}^{n}$ (with $\{x_i\} \in K, \forall i \in \{0,...,k\}$) such that
\begin{itemize}
    \item[(1)] if $\sigma \in K$ and $\kappa \subset \sigma$, then $\kappa \in K$ 
    \item[(2)] if $\sigma_1, \sigma_2 \in K$, then $\sigma_{1 \cap 2} = \sigma_1 \cap \sigma_2 \in K$
\end{itemize}

\textit{Example 2.} In Example 1, given $X=\{1,2,3\}$, a simplicial complex $K$ can be $\{\{1\},\{2\},\{3\},\{1,2\},$ $\{2,3\},\{1,3\},\{1,2,3\}\}$. For $\sigma=[1,2,3]$ and $\kappa=\{1,2\}$, we have $\kappa \in K$. The intersection between $\{1,2\}$ and 
$\{2,3\}$ is $\{2\} \in K$.

There are different types of simplicial complexes including Vietoris-Rips complex and $\tilde{C}$ech complex. Given a metric space $(X, d_{X})$ and a real number $\varepsilon > 0$, the $\tilde{C}$ech complex denoted $\tilde{\mathcal{C}}_{\varepsilon}(X)$ is a set of simplices $\sigma = [x_0,...,x_k]$ such that for each $\sigma \subseteq X$ and $\sigma \in \tilde{\mathcal{C}}_{\varepsilon}(X)$, the $\varepsilon$-balls $B(x_i, \varepsilon)$, centered at $x_i \in \sigma$, has a non-empty intersection, i.e.,
\[\bigcap_{i=0}^{k} B(x_i, \varepsilon) \neq \emptyset
\]

with $B(x, \varepsilon) = \{ x' \in X : d_X(x, x') \leq \varepsilon\}$.  A set of simplices $\sigma = [x_0,...,x_k]$ such that $d_{X}(x_i, x_j) \leq \varepsilon$, for any $x_i, x_j \in \sigma$ is called the Vietoris-Rips complex and it is denoted $\mathcal{R}_{\varepsilon}(X)$. As we can see, Vietoris-Rips and $\tilde{C}$ech complexes are closely related. Both complexes have a natural inclusion, meaning that, given $\varepsilon, \varepsilon' > 0$, such that $\varepsilon \leq \varepsilon'$, we have $\tilde{\mathcal{C}}_{\varepsilon}(X) \subseteq \tilde{\mathcal{C}}_{\varepsilon'}(X)$ (resp. $\mathcal{R}_{\varepsilon}(X) \subseteq \mathcal{R}_{\varepsilon'}(X)$). In~\cite{de2007coverage}, it is proven that there is a chain of inclusions between $\mathcal{R}_{\varepsilon}(X)$ and $\tilde{\mathcal{C}}_{\varepsilon}(X)$ (by considering that the radius is $\varepsilon/2$) such that
\[
   \mathcal{R}_{\varepsilon'}(X) \subset \tilde{\mathcal{C}}_{\varepsilon}(X) \subset \mathcal{R}_{\varepsilon}(X), \text{with} \; \dfrac{\varepsilon}{\varepsilon'} \leq \sqrt{\dfrac{2n}{n+1}}
\]
The Vietoris-Rips complex will be used later to define persistence diagrams. The $\tilde{C}$ech complex is often associated to covers to form a nerve~\cite{hatcher2005algebraic}. $\tilde{C}$ech complex is the foundation of the TDA Mapper algorithm, that will be defined later. 

\textbf{Nerve.} Let $\mathcal{U}=\{U_{i}\}_{i \in I}$ be a finite covering of $X$ and $I=\{i_0,i_1,..,i_{|I|}\}$ is an indexing set. The nerve of $\mathcal{U}$ is the simplicial complex $\tilde{\mathcal{C}}(\mathcal{U})$ indexed in $I$ such that the $k$-simplex $\sigma' = [U_{i_0},...,U_{i_k}] \in \tilde{\mathcal{C}}(\mathcal{U})$ if and only if 
\[\bigcap_{p=0}^{k} U_{i_p} \neq \emptyset
\]

$\mathcal{U}$ is an open covering of $X$ if and only if for each member $U_{i}$, $U_{i}$ is an open set, i.e., $U_{i}$ is contained in the topology of $X$. 

\textit{Example 3.} Given $X=\{1,2,3\}$, $\mathcal{U} = \{\{1\},\{2\},\{3\},\{1,2\},\{2,3\},\{1,3\},\{1,2,3\}\}$ is an open cover. The members of $\mathcal{U}$ (e.g., $\{2,3\}, \{1,2,3\}$) are open sets; since they are contained in the topology $\{\emptyset\} \cup \mathcal{U}$ of $X$. 

Let $f: X \rightarrow Y$ be a continuous mapping function such that $Y$ is a space with a covering $\mathcal{U}=\{U_{i}\}_{i \in I}$, the inverse $f^{-1}(U_{i})$ is also an open covering of $X$.

\textbf{Homology.} Given a simplicial complex $K$, let $C_*(K)$ be a chain complex encoding information of $K$. A chain complex is a sequence of abelian groups $C_0(K),C_1(K),..., C_{j-1}(K), C_{j}(K),...$ connected by boundary maps $\phi_j: C_{j} \rightarrow C_{j-1}$, such that
\[
  ... \xrightarrow[]{\phi_{j+1}} C_{j}(K) \xrightarrow[]{\phi_{j}} C_{j-1}(K)\xrightarrow[]{\phi_{j-1}} ... \xrightarrow[]{\phi_{1}} C_0(K) \xrightarrow[]{\phi_{0}} \textsf{0}
\]

where $\textsf{0}$ is the zero or trivial group and for every $j$, $\phi_j \circ \phi_{j+1} = \textsf{0}$. Therefore, the kernel of $\phi_j$ denoted $\textsf{ker}(\phi_j)$ and the image $\textsf{im}(\phi_j)$ are defined such that $\textsf{im}(\phi_j) \subseteq \textsf{ker}(\phi_j)$. Then, the quotient group $H_j$ can be constructed such that 
\[
   H_j(K) = \textsf{ker}(\phi_j)/\textsf{im}(\phi_j)
\]
and denotes the $j$-th simplicial homology group of $K$. $H_j(K)$ represents the number of holes in $K$ with a $j$-th dimensional boundary. The rank of $H_j(K)$ represents the $j$-th Betti number $\beta_j(K)$ of $K$ given by,
\[
\beta_j(K) = \textsf{rank}(H_j(K)) = \textsf{dim}(\textsf{ker}(\phi_j))-\textsf{dim}(\textsf{im}(\phi_j))
\]

\subsection{TDA Mapper}

Given a data set $X$, TDA Mapper~\cite{singh2007topological} summarizes data points in $X$ through the nerve of the collection of connected components of the open sets $\{f^{-1}(U_{i})\}_{i \in I}$, where $f: X \rightarrow \mathbb{R}^{n}$ is a continuous filter/lens function, $U_{i} \in \mathcal{U}$ (for each $i$ in the indexing set $I$), and $f^{-1}(U_{i})$ are preimages. The nerve is a graph consisting of nodes and edges that retain/extract important geometric features of the original data (e.g., holes, components, branches). The TDA Mapper graph is denoted $\mathcal{M}(X, \mathcal{U}, f)$ and it is computed as follows:
\begin{enumerate}
    \item[1.] Map the dataset $X$ into Euclidean space $\mathbb{R}^{n}$ using the filter/lens function $f: X \rightarrow \mathbb{R}^{n}$
    \item[2.] Cover the image $f(X)$ of the dataset $X$ in $\mathbb{R}^{n}$, by the collection of open sets $\mathcal{U}=\{U_{i}\}_{i \in I}$
    \item[3.] For each preimage $f^{-1}(U_{i})$, apply a clustering algorithm (e.g., k-nearest neighbor, density-based) resulting into $i_p$ clusters $C_{i_1},...,C_{i_p}$ in the $i$-th preimage
    \item[4.] Build the nerve from the set of all clusters $\{C_{i_1},...,C_{i_p}\}$ with all vertices $v_{i_k} \in C_{i_k}$, edges between $v_{j_k}$ and $v_{k_l}$ are added if and only if $C_{j_k} \cap C_{k_l} \neq \emptyset$
\end{enumerate}

The TDA Mapper graph has various parameters that must be carefully chosen. For example, the filter/lens function depends of the data features and the clustering algorithm must be provided to compute preimages. The computation of TDA Mapper algorithm is polynomial but can be parallelized while keeping the correctness of results~\cite{hajij2020parallel}.

\subsection{Persistence Homology}

Persistence homology~\cite{edelsbrunner2000topological, zomorodian2005computing, ghrist2008barcodes, chazal2009gromov, adams2017persistence, carriere2021optimizing} encodes topological information in the data set $X$ through a filtration of the simplicial complex $K$ of $X$. The filtration of $K$ is a nested sequence of subcomplexes $\{K_i\}_{i \in I}$ that starts with an empty complex $K_{i_0}=\emptyset$ and ends with the complete complex $K_{i_m}=K (m = |K|)$, i.e.,
\[
  \emptyset = K_{i_0} \subseteq K_{i_1} \subseteq ... \subseteq K_{i_m} = K
\]

where $i_0 \leq i_1 \leq ... \leq i_m$ and $K = \cup_{i \in I} K_i$. In a specific case, $K$ can be complexes such as Rips-Vietoris or $\tilde{C}$ech. Then, the family of Rips-Vietoris complexes $\{\mathcal{R}_{i}(X)\}_{i \in I}$ and $\tilde{C}$ech complexes $\{\tilde{\mathcal{C}}_{i}(X)\}_{i \in I}$ are filtrations, where $i$ represents the resolution at which dataset $X$ is considered. 

\textbf{Persistence diagrams.} The inclusion maps between the previous subcomplexes $K_i$ induces a chain of $k$-th dimensional homology groups (also called persistent modules), connected by homomorphism maps $\PHI^r_{s}: H_k(K_{i_s}) \rightarrow H_k(K_{i_r})$ with $0 \leq s < r \leq +\infty$, such that
\begin{multline*}
  H_k(K_{i_0}) \xrightarrow[]{\PHI^1_0} H_k(K_{i_1}) \xrightarrow[]{\PHI^2_1} ... \xrightarrow[]{\PHI^s_{s-1}} H_k(K_{i_s}) \xrightarrow[]{\PHI^r_s} \\
  H_k(K_{i_r}) \xrightarrow[]{\PHI^{r+1}_r} ... \xrightarrow[]{\PHI^m_{m-1}} H_k(K_{i_m})
\end{multline*}

where $\PHI^r_s =\PHI^r_{r-1} \circ ... \circ \PHI^{s+1}_s$. These persistent modules are encoded in a multi-set of points $\textsf{\small Dgm}_k K$, called $k$-th persistence diagram of $K$. $\textsf{\small Dgm}_k K$ is the set $\{(b_i, d_i) \in \mathbb{R}^2\;|\;i \in I, b_i < d_i\}$, where $b_i$ is the birth where the topological property $(b_i, d_i)$ appears and $d_i$ where it disappears. The persistence concept extends to the case of sublevel sets filtration $\{K_r\}_{r \in \mathbb{R}}$ of the continuous function $f: X \rightarrow \mathbb{R}$ under a tameness condition~\cite{chazal2009proximity}, where data points in $X$ are vertices of the simplicial complex $K$ and the family of subcomplexes $K_r = \{\sigma \in K : f(\sigma) \leq \rho_r\}$ w.r.t the inclusion (i.e., $K_r \subseteq K_{r'}$ if $r \leq r'$). The threshold $\rho_r$ allows one to track topological changes (e.g., appearing, merged, filled, and disappearing components) in the filtration as its values vary between $-\infty$ and $+\infty$. 

\textbf{Stability.} Persistence diagrams have the advantage to be stable under small perturbations~\cite{cohen2007stability}. Given two continuous tame functions $f, g: X \rightarrow \mathbb{R}$, we can induce two persistence diagrams based on their sublevel sets filtration $\textsf{\small Dgm}_k f$ and $\textsf{\small Dgm}_k g$. Given a bijection $\gamma$ between the diagrams, the distance $d_b$ given by
\[
   d_b(\textsf{\small Dgm}_k f,\textsf{\small Dgm}_k g) = \operatorname*{\text{inf}}_{\gamma} \operatorname*{\text{sup}}_{x \in (\textsf{\footnotesize Dgm}_k f) \cup \Delta} ||x-\gamma(x)||_{\infty} 
\]

is called the bottleneck distance and it is computed by taking the infimum over all supremum $L_{\infty}$-distance between matched points in the two diagrams, where the diagonal set $\Delta = \{(\alpha, \alpha)\;|\;\alpha \in \mathbb{R}\}$, $\gamma$ ranges over all bijective map between the multi-sets $(\textsf{\small Dgm}_k f) \cup \Delta$ and $(\textsf{\small Dgm}_k g) \cup \Delta$, and the $L_{\infty}$-distance $||.||_\infty$ is defined such that for any $x=(x_0,...,x_n) \in X$, $||x||_{\infty} = \max \{|x_0|,..,|x_n|\}$. Other distances such as $p$-Wasserstein~\cite{cohen2010lipschitz} and Gromov-Hausdorff~\cite{chazal2009gromov} are also used. Let us consider the distance between functions $||f-g||_{\infty} = \operatorname*{\text{sup}}_{x \in X} |f(x)-g(x)|$. In~\cite{cohen2007stability}, it is proven that the bottleneck distance is bounded, i.e.,
\[
   d_b(\textsf{\small Dgm}_k f,\textsf{\small Dgm}_k g) \leq  ||f-g||_{\infty} 
\]
thus showing the stability of persistence diagrams. 

\textbf{Persistence features.} In order to make persistence features consumable by machine learning classifiers, several methods have been defined such as kernels~\cite{reininghaus2015stable,kusano2017kernel}, persistent landscapes~\cite{bubenik2015statistical}, persistent entropy~\cite{atienza2019persistent, atienza2020stability}, and persistent images~\cite{adams2017persistence}. 



The computation of persistence diagrams is at most cubic but can be parallelized in a distributed memory environment~\cite{bauer2014distributed}. 

\begin{figure*}[h!]
\centering
\subfloat[CIC-MalMem-2022]{\includegraphics[scale=0.67]{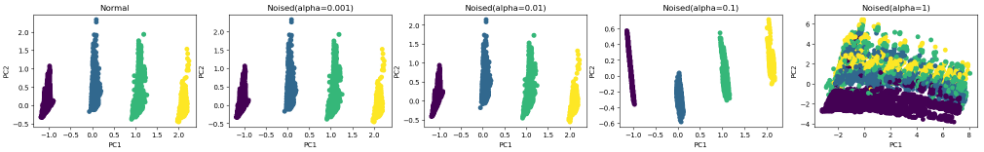}\label{fig1}}\\
\subfloat[CCCS-CIC-AndMal-2020]{\includegraphics[scale=0.75]{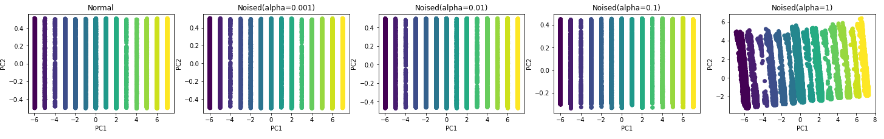}\label{fig2}}
\caption{PCA analysis}
\end{figure*}

\subsection{Tomato}

Topological Mode Analysis Tool (Tomato)~\cite{chazal2013persistence} takes a graph built from the data set $X$ where vertices $x_i$ are data points and edges $(x_i, x_j)$ are pair of connected data points, and assigns each vertex $x_i$ to a non-negative value $\tilde{f}(x_i)$, where $\tilde{f}: X \rightarrow \mathbb{R}^{+}$ is a density estimator of the density function $f: X \rightarrow \mathbb{R}$. The persistence diagram of $\tilde{f}$ is computed to determine the prominences of its peaks (i.e., highest $\tilde{f}$ values) allowing to identify peaks of the true density function $f$. The set of $\tilde{f}$-value peaks form a spanning forest where each tree forms an ascending region of a peak of $f$. 

Next, Tomato merges trees by iterating over the vertices while maintaining a union-find data structure $\mathcal{U}$, where each entry $e$ corresponds to a union of trees of the spanning forest. If vertex $x_i$ is a peak of $\tilde{f}$, a new entry $e$ is created in $\mathcal{U}$ and added to a root map $r$ (i.e., $r(e) = x_i$, where $r(e)$ is the root of entry $e$). Otherwise, entry $e_i$ already exists in $\mathcal{U}$. Thus, a set of entries of $\mathcal{U}$ containing neighbors of $x_i$ is computed and for each entry $e$ in the set, if $e, e_i$ are different and one of them have at least $\delta$-prominent root, i.e.,
\[
  \textsf{min}\{\tilde{f}(r(e)), \tilde{f}(r(e_i))\} < \tilde{f}(x_i) + \delta,\;\text{with}\; e\neq e_i
\]
then, $e_i \cup e$ is added to $\mathcal{U}$, i.e., 
\[
 r(e \cup e_i) = \argmax_{r(e), r(e_i)} \tilde{f}
\]
At the end, the root map $r$ contains the $\delta$-prominent peaks of $\tilde{f}$ and Tomato returns a subset of the collection of clusters stores in entries $e$ of $\mathcal{U}$ such that $\tilde{f}(r(e)) \geq \delta$. The computation of Tomato is almost-linear running time in the size of the graph built on the input dataset and a linear memory usage in the number of data points~\cite{chazal2013persistence}.

\subsection{TDA tools and performance}

Several tools have been proposed such as Ripser, Scikit-TDA, Gudhi, Dionysus, JavaPlex, Perseus, JHoles, DIPHA, and PHAT. In~\cite{otter2017roadmap}, a performance comparison have been done on a (1) Dell Sandybridge cluster with 1728 cores of 2 GHz Xeon Sandybridge, 64 GB RAM in 80 nodes and 128 GB RAM in 4 nodes, and 20 TB disk; and (2) a share-memory IBM system x3550 M4 server with 16 cores of 3.3 GHz, 768 GB RAM, and 3 TB disk. Ripser, Gudhi, and DIPHA had the best performance on average but for cubical complexes, Gudhi outperformed DIPHA. Ripser and Gudhi are more mature than Scikitlearn-TDA. Thus, we use Ripser and Gudhi tools to run experiments in this work.

%

\section{Methodology}\label{methodology}

The \textbf{goal} of this study is to show how TDA techniques (TDA Mapper, Persistence Homology, Tomato) can be used to better analyze complex malware relationships and better detect new malware samples (no noise, under noise). The \textbf{perspective} of this study is that of malware analysts and researchers can exploit TDA techniques to improve malware analysis and also use them for malware detection along the existing methods (e.g., random forest, xgboost, lightgbm). The process used for malware analysis and detection consists in four steps: (i) collect datasets based on reputed malwares from Desktop and mobile platforms (e.g., Android, Windows) since they are more targeted in real-world, (ii) analyze and extract features based on existing methods (PCA, t-SNE, UMAP) and TDA methods (TDA Mapper, Persistence Homology, Tomato), (iii) detect new malware samples (no noise, under noise) based on features, and (iv) evaluate the performance of the models (detection rate, false positive rate, f1-score, execution time, memory usage). The \textbf{context} of this study consists of 
58 596 records of malware memory samples with 29 298 benign and 29 298 malicious from the Canadian Institute for Cybersecurity in 2022 (CIC-MalMem-2022) \footnote{https://www.unb.ca/cic/datasets/malmem-2022.html} and 400 000 Android static/dynamic malware samples with 200 000 benign and 200 000 malicious from the Canadian Centre for Cyber Security (CCCS) in collaboration with the Canadian Institute for Cybersecurity in 2020 (CCCS-CIC-AndMal-2020)~\footnote{https://www.unb.ca/cic/datasets/andmal2020.html}. 
To achieve our goal, we address the following research questions:
\begin{quote}
	\textbf{RQ$_1$:} \textit{What are the best feature engineering techniques for malware analysis ?} The aim is to identify methods among TDA techniques (i.e., TDA Mapper, Persistence Homology, Tomato) and existing techniques (i.e., PCA, UMAP, t-SNE) that allow one to precisely and quickly analyze complex malware relationships.
\end{quote}

\begin{quote}
	\textbf{RQ$_2$:} \textit{What are the most performant techniques for malware detection ?} This research question helps to identify techniques having both the best detection rate, the lowest false positive rate, the lowest execution time, and the lowest memory usage. 
\end{quote}

\begin{quote}
	\textbf{RQ$_3$:} \textit{What are the most robust techniques against noisy input data ?} The goal is to identify techniques that achieve the best detection rate and the lowest false positive rate with no noise as well as under noise. 
\end{quote}

To conclude our research questions, some recommendations will be proposed to cyber analysts to deploy the best-identified models for malware detection at scale.

\begin{figure*}[]
\centering
\subfloat[CIC-MalMem-2022]{
\includegraphics[scale=0.64]{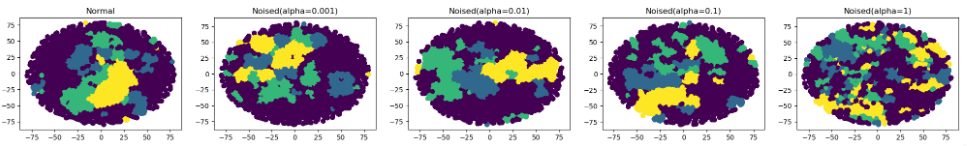}
\label{fig3}
}\\
\subfloat[CCCS-CIC-AndMal-2020]{
\includegraphics[scale=0.63]{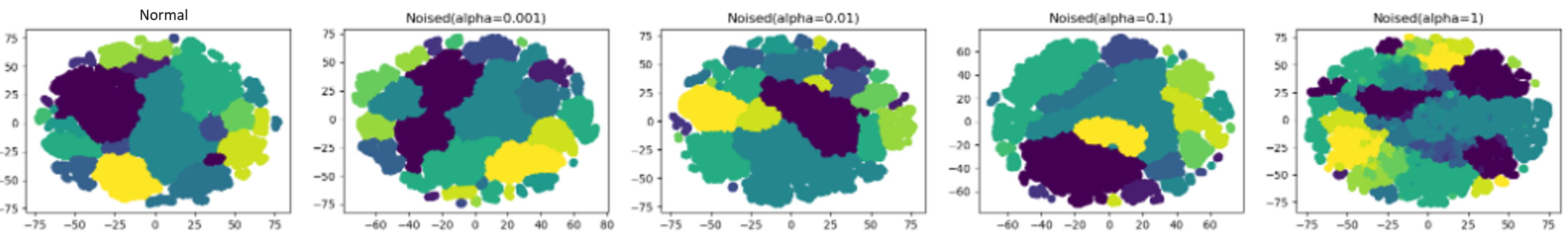}
\label{fig4}
}
\caption{t-SNE Analysis}
\end{figure*}

\begin{figure*}[]
\centering
\subfloat[CIC-MalMem-2022]{
\includegraphics[scale=0.63]{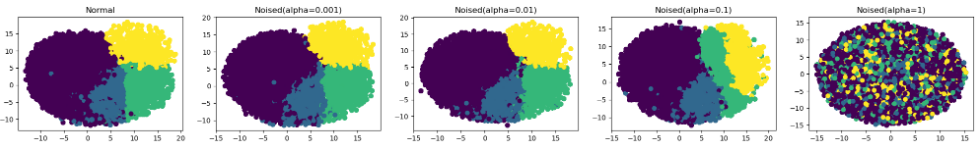}
\label{fig5}
}\\
\subfloat[CCCS-CIC-AndMal-2020]{
\includegraphics[scale=0.63]{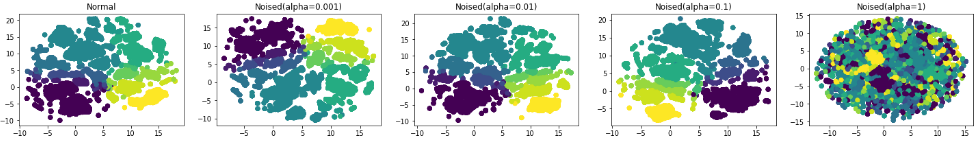}
\label{fig6}
}
\caption{UMAP Analysis}
\end{figure*}

\subsection{Data collection}

We have selected CIC-MalMem-2022 
and CCCS-CIC-AndMal-2020 datasets based on their reputation, diversity, and recency. 
A detailed description of the datasets is  provided in the following paragraphs. 

\subsubsection{CIC-MalMem-2022}

CIC-MalMem-2022 dataset consists of 58 596 records of malware memory samples with 29 298 benign and 29 298 malicious. Malware samples are grouped in 3 categories and 15 families in total captured in 2022: \textit{Ransomware} such as Maze, Shade, Ako, and Pysa; \textit{Spyware} such as 180Solutions, Coolwebsearch, Gator, Transponder, and TIBS; and \textit{Trojan-Horse} such as Reconyc, Zeus, Emotet, Refroso, and Scar. The dataset has the following features extracted using Volatility tool: process list, DLL list, handles, LDR modules, malfind, psxview, svcscan, and callbacks. The dataset has also 4 classes: Benign, Ransomware, Spyware, and Trojan Horse.

\subsubsection{CCCS-CIC-AndMal-2020}

CCCS-CIC-AndMal-2020 dataset consists of 400 000 Android static and dynamic malware samples with 200 000 benign and 200 000 malicious. Malware samples are grouped in 14 categories and 191 families in total captured in 2020, including  \textit{Adware}; \textit{Backdoor}; \textit{File Infector}; \textit{Ransomware}; \textit{Riskware}; \textit{Scareware}; \textit{Trojan}; \textit{Trojan-Banker}; \textit{Trojan-Dropper}; \textit{Trojan-SMS}; \textit{Trojan-Spy}; and \textit{zero-day}. The dataset has static and dynamic features. Static features include activities, services, broadcast receivers/providers, intent actions, intent categories, permissions requested by application, metadata, and system features. Dynamic features include memory, API, network, battery, logcat, and process. The dataset has also 14 classes similar to the aforementioned categories. Unlike CIC-MalMem-2022, CCCS-CIC-AndMal-2020 is imbalanced i.e., some malware classes have more examples than others. Thus, it will help us to see how TDA techniques perform in this case compared to existing techniques (i.e., PCA, UMAP, t-SNE).

\subsubsection{Noised CIC-MalMem-2022 and CCCS-CIC-AndMal-2020}

A noise is added to datasets to evaluate the robustness of the techniques. The noise is a random gaussian normal function $p : \mathbb{R} \rightarrow \mathbb{R}$ with the mean $\mu$ and the standard deviation $\sigma$. Given a data $x_i \in X$ (dataset), the noised dataset $\tilde{X} = (\tilde{x}_i)_{i \in \{0,...,|X|-1\}}$ is given by

\[
\begin{cases}
  \tilde{x_i} = x_i + \alpha *p(x_i) \text{, where } \alpha \in [0,1]\\
   p(x_i) = \dfrac{1}{\sqrt{2 \pi \sigma^2}} \exp{-\dfrac{(x_i-\mu)^2}{2\sigma^2}}
\end{cases}
\]

In this work, we consider the mean $\mu = 0$ and the standard deviation $\sigma = 0.1$ to generate noised datasets. For each dataset, we generate 4 noised datasets when $\alpha$ has value 0.001, 0.01, 0.1, and 1. This approach for noising datasets is similar to the additive gaussian noise extensively used in the past decades for image processing~\cite{1055184, 6305478}. 

\begin{figure*}[]
\centering
\subfloat[CIC-MalMem-2022]{
\includegraphics[scale=0.64]{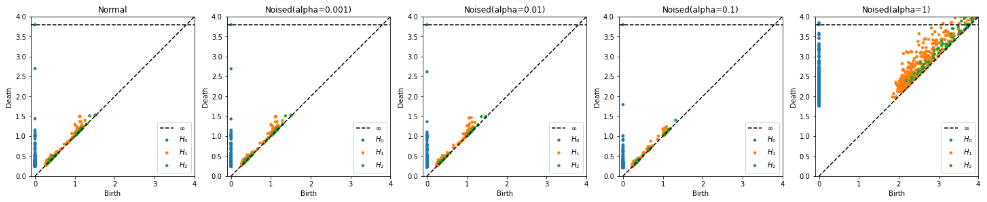}
\label{fig15}
}\\
\subfloat[CCCS-CIC-AndMal-2020]{
\includegraphics[scale=0.64]{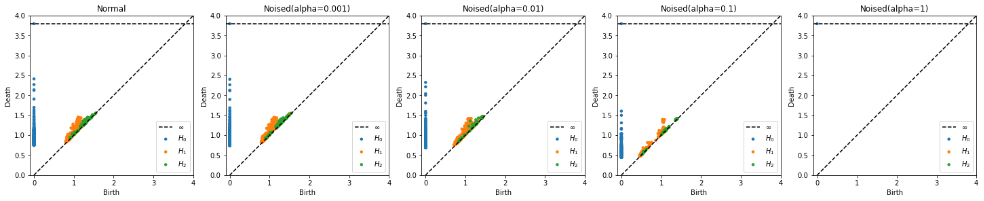}
\label{fig13}
}
\caption{Persistence Diagrams}
\end{figure*}

\begin{figure*}[]
\centering
\subfloat[CIC-MalMem-2022]{
\includegraphics[scale=0.53]{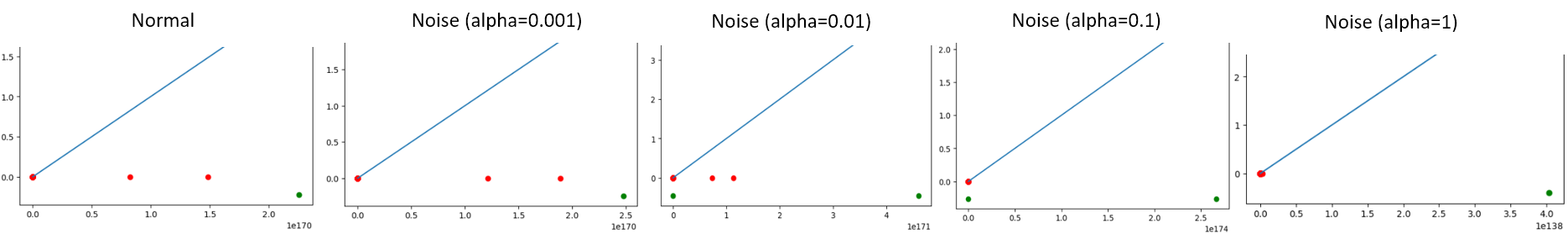}
\label{fig18}
}\\
\subfloat[CCCS-CIC-AndMal-2020]{
\includegraphics[scale=0.53]{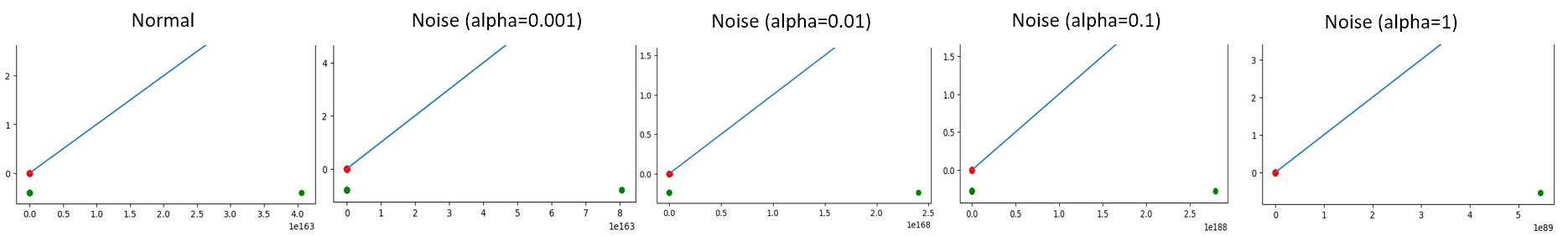}
\label{fig17}
}
\caption{Tomato}
\end{figure*}

\begin{figure*}[]
\centering
\subfloat[PCA TDA Mapper]{
\includegraphics[scale=0.51]{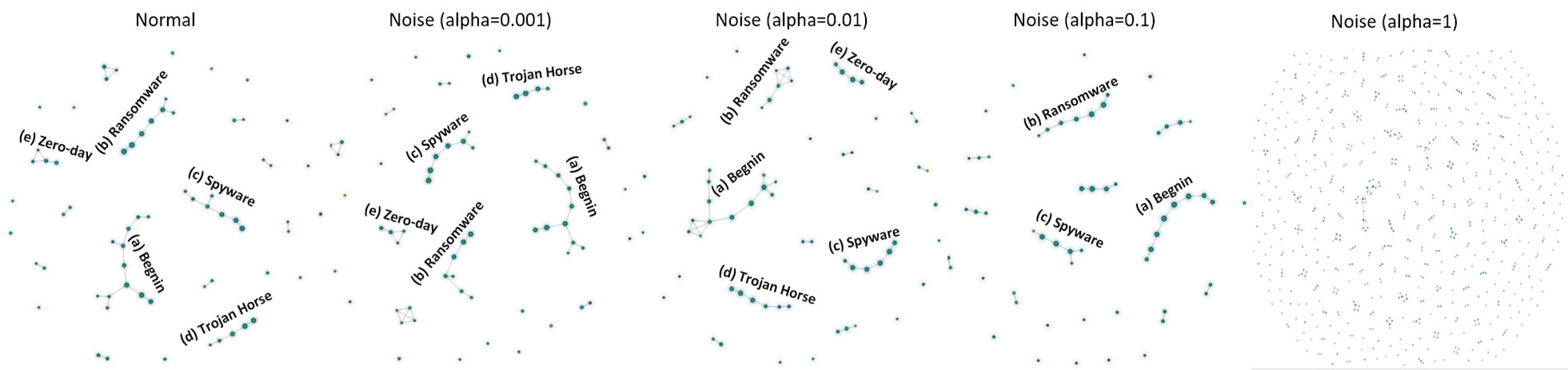} 
\label{fig7}
}\\
\subfloat[UMAP TDA Mapper]{
\includegraphics[scale=0.51]{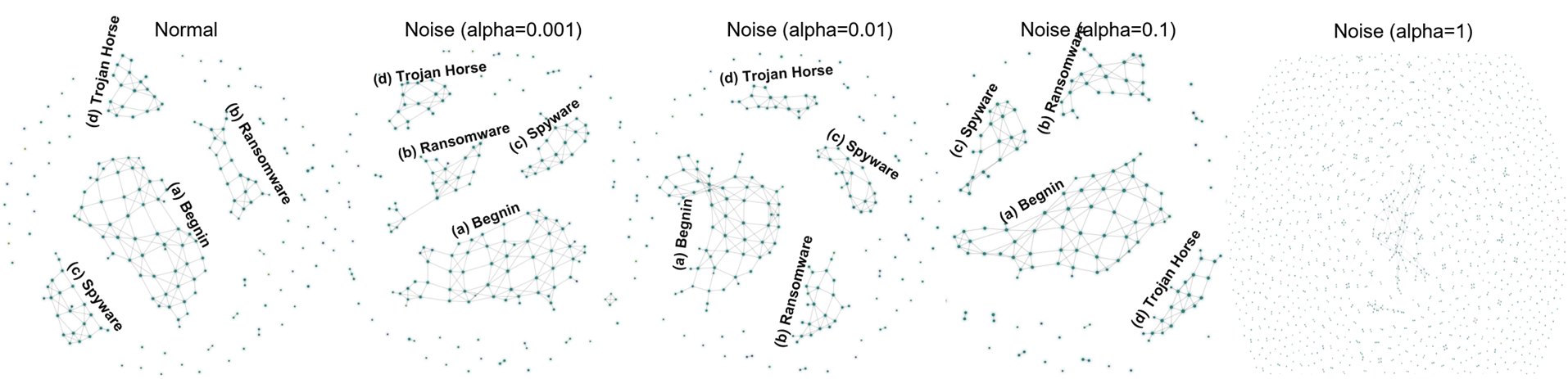}
\label{fig8}
}\\
\subfloat[t-SNE TDA Mapper]{
\includegraphics[scale=0.51]{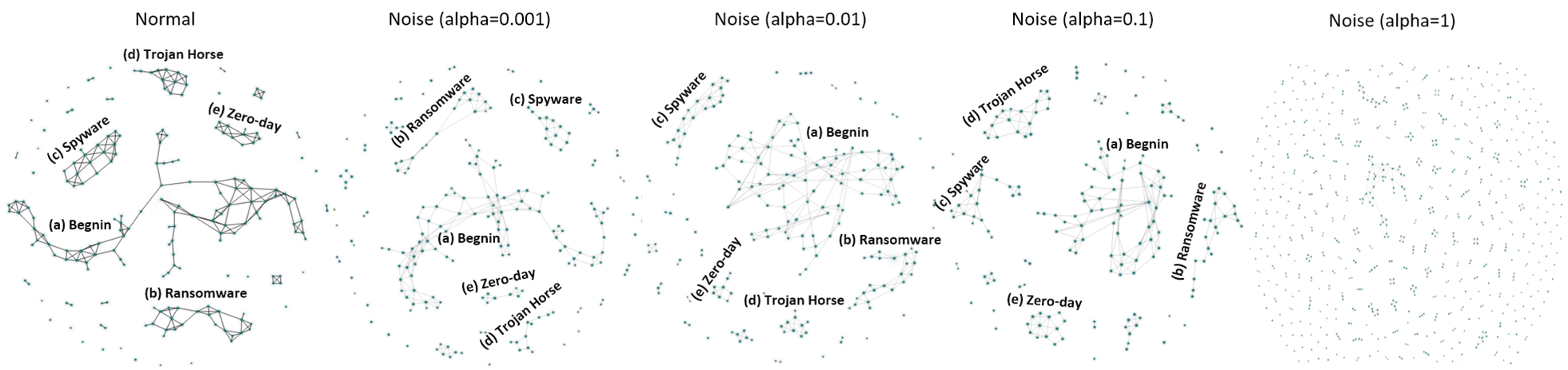}
\label{fig9}
}
\caption{TDA Mapper algorithm on the CIC-MalMem-2022 dataset}
\end{figure*}

\subsection{Data analysis}

In order to answer RQ1, we extracted features from malware datasets using existing techniques (e.g., PCA, t-SNE, UMAP) and TDA techniques (e.g., TDA Mapper, Persistence Homology, Tomato). This section describes our feature extraction and selection process. The features are selected using the following criteria~\cite{rand1971objective, van1993categories}: high density of connected points, uniqueness of clusters, cluster topology (e.g., linear, convex), small within-cluster dissimilarity, large between-cluster dissimilarity, execution time, and memory usage. 
For each dataset, we remove label classes (i.e., \textit{Class} column for CIC-MalMem-2022, and \textit{Category} column for CCCS-CIC-AndMal-2020) and perform label encoding and min-max scaling before executing the analysis techniques. We have conducted our experiments on a Dell Precision 5530 with 32 GB RAM, 2 TB SSD, and a processor of 2.6 GHz speed and 12 CPUs.

\subsubsection{PCA}

Fig.~\ref{fig1} shows features from malware samples in the CIC-MalMem-2022 dataset using PCA. For PCA, we have used the scikit-learn library and set the number of components to 2. PCA identified 4 distinct features proportional to the 4 classes: Trojan Horse, Benign, Ransomware, and Spyware. From the CCCS-CIC-AndMal-2020, PCA extracted the following classes in normal case (see Fig.~\ref{fig2}): Adware, Backdoor, No\_Category, PUA, Ransomware, Riskware, Scareware, Trojan\_Banker, Trojan\_Dropper, Trojan\_SMS, Trojan\_Spy, and Zero\_Day.

\subsubsection{t-SNE}

Based on CIC-MalMem-2022 dataset, Fig.~\ref{fig3} shows features from malware samples using t-SNE. For t-SNE, we have used the scikit-learn library and set the number of components to 2 and the perplexity to 20 after hyperparameter tuning (see Section~\ref{classif}). t-SNE identified 4 compact features equivalent to the 4 labeled classes: Benign, Ransomware, Spyware, and Trojan Horse. For CCCS-CIC-AndMal-2020, t-SNE also extracted the following features in normal case (see Fig.~\ref{fig4}): Adware, Backdoor, No\_Category, PUA, Ransomware, Riskware, Scareware, Trojan\_Banker, Trojan\_Dropper, Trojan\_SMS, Trojan\_Spy, and Zero\_Day.

\subsubsection{UMAP}

Fig.~\ref{fig5} shows features from malware samples in the CIC-MalMem-2022 dataset using UMAP. Using the UMAP function of python, we have set the number of components to 2, the number of neighbors to 5, and the metric norm to euclidean. Then, UMAP identified 4 compact features proportional to the 4 classes: Benign, Ransomware, Spyware, and Trojan Horse. For CCCS-CIC-AndMal-2020, UMAP also extracted the following classes in normal case (see Fig.~\ref{fig6}): Adware, No\_Category, PUA, Ransomware, Riskware, Scareware, Trojan\_Dropper, Trojan\_SMS, Trojan\_Spy, and Zero\_Day.

\subsubsection{TDA Mapper}

As stated earlier, TDA Mapper algorithm needs a lens function (typically a clustering algorithm) in order to create nerves from data. We have computed the TDA Mapper algorithm using Gudhi Library and used different lens function such as PCA, UMAP, and t-SNE. TDA Mapper is fully customizable and it allows an automatic multi-resolution glimpse into the structure and information behind the data generation process~\cite{10.5555/3491440.3492052}.

Based on the CIC-MalMem-2022 dataset, Fig.~\ref{fig7} shows features from the TDA Mapper algorithm with PCA as lens. PCA TDA Mapper identified 4 significant nerve features similar to the 4 classes: (a) Benign, (b) Ransomware, (c) Spyware, and (d) Trojan Horse. In addition, PCA TDA Mapper identified a new malware class (e) called Zero-day. In practice, data from the Zero-day class are used to retrain the models in order to identify new types of malwares. 
From the CIC-MalMem-2022 dataset, Fig.~\ref{fig8} describes features from TDA Mapper with UMAP as lens. UMAP TDA Mapper also identified 4 significant nerve features: (a) Benign, (b) Ransomware, (c) Spyware, and (d) Trojan Horse. Like PCA TDA Mapper, t-SNE TDA Mapper identified 5 representative nerve features (see Fig.~\ref{fig9}): (a) Benign, (b) Ransomware, (c) Spyware, (d) Trojan Horse, and (e) Zero-day.

On the CCCS-CIC-AndMal-2020, PCA TDA Mapper identified the following features (see Fig.~\ref{fig10}) in normal case: (a) Riskware, (b) Adware, (c) Zero\_Day, (d) Ransomware, (e) Trojan-Spy, (f) Trojan-SMS, (g) Trojan-Dropper, (h) No Category, (i) PUA, (j) Scareware, (k) Backdoor, and (l) Trojan-Banker. In Fig.~\ref{fig11}, UMAP TDA Mapper also identified the following features in normal case: (a) Riskware, (b) Adware, (c) Zero\_Day, (d) Ransomware, (e) Trojan-Spy, (f) Trojan-SMS, (h) No Category, (i) PUA, and (j) Scareware. In Fig.~\ref{fig12}, 14 significant features are identified by t-SNE TDA Mapper in normal case including (a) Riskware, (b) Adware, (c) Zero\_Day, (d) Ransomware, (e) Trojan-Spy, (f) Trojan-SMS, (h) No Category, (i) PUA, (j) Scareware, and (k) Backdoor.

\subsubsection{Persistence Homology}

To build persistence diagram, we compute the Vietoris-Rips complex using the Ripser library and we have set the dimension to 2, the metric is euclidean, and the number of points to subsample for a faster computation. Then, we obtain birth-death coordinates represented by three homology groups: H0, H1, and H2. Fig.~\ref{fig13} and Fig.~\ref{fig15} respectively show the persistence diagrams of the CCCS-CIC-AndMal-2020 and CIC-MalMem-2022 datasets.



\subsubsection{Tomato}

Tomato (Topological Mode Analysis Tool) clustering is computed using the Gudhi library. We set the default density type to Document-Term Matrix (DTM) and the number of neighbors for a K-nearest neighbors (KNN) graph to 100. Then, we compute the fitness function on data. Fig.~\ref{fig17} and Fig.~\ref{fig18} respectively show the cluster points computed using Tomato on CCCS-CIC-AndMal-2020 and CIC-MalMem-2022 datasets. 
\begin{figure*}[]
\centering
\subfloat[PCA TDA Mapper]{
\includegraphics[scale=0.51]{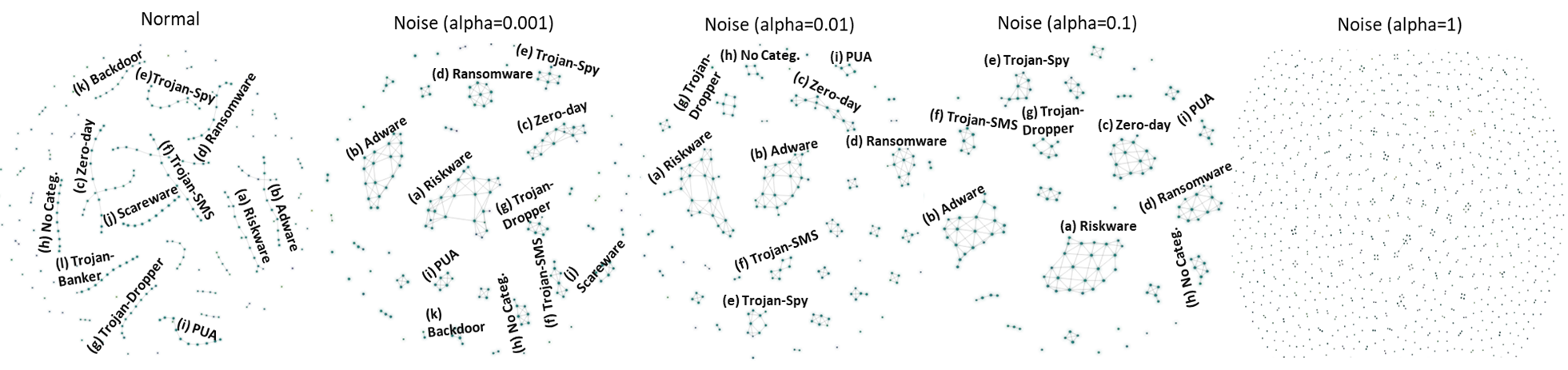}
\label{fig10}
}\\
\subfloat[UMAP TDA Mapper]{
\includegraphics[scale=0.51]{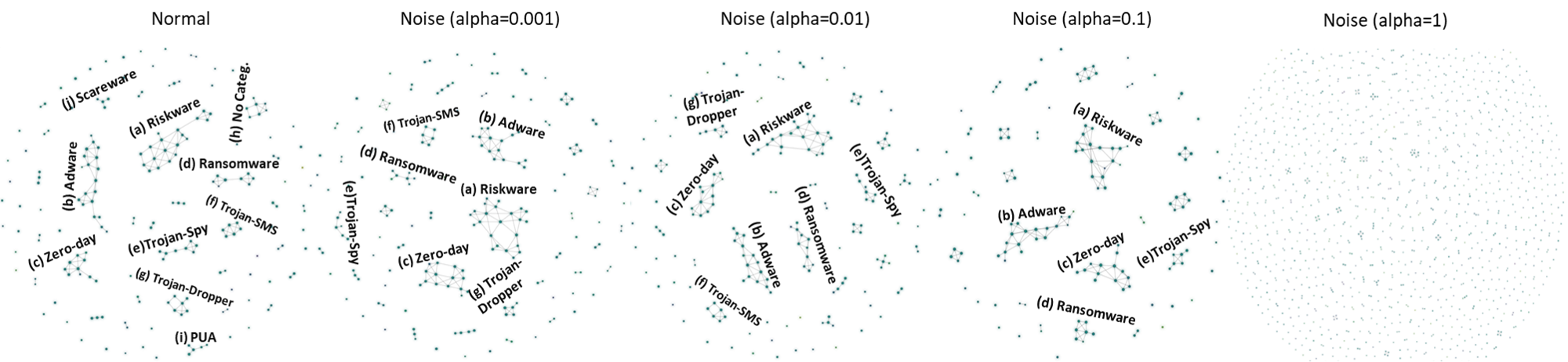}
\label{fig11}
}\\
\subfloat[t-SNE TDA Mapper]{
\includegraphics[scale=0.51]{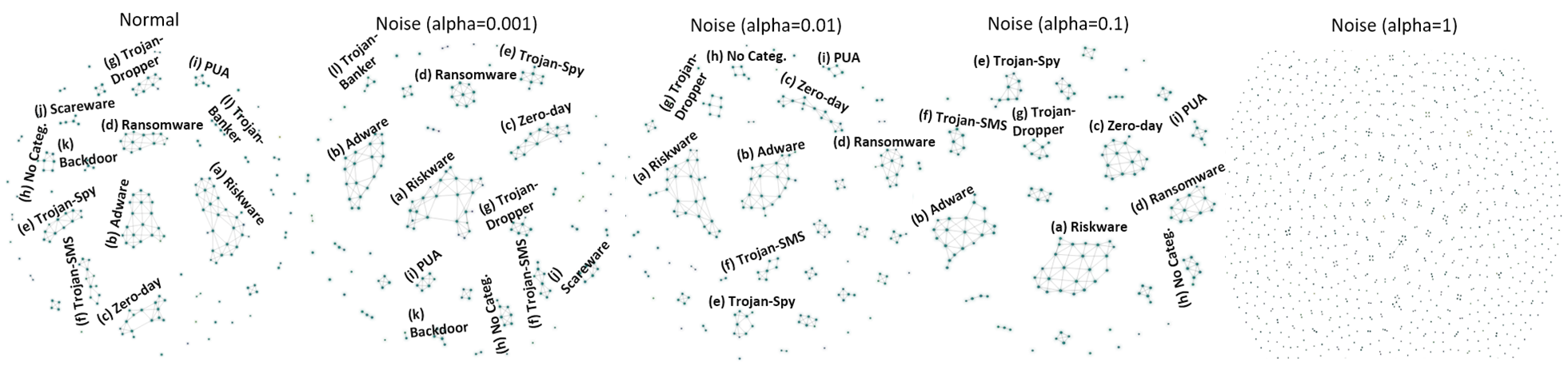}
\label{fig12}
}
\caption{TDA Mapper algorithm on the CCCS-CIC-AndMal-2020}
\end{figure*}

\begin{figure*}[]
\centering
\subfloat[PCA classification]{
\includegraphics[scale=0.52]{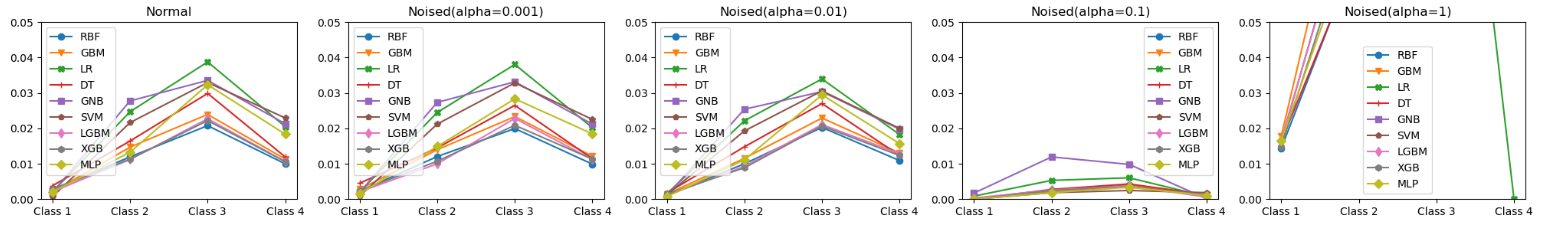}
\label{fig30}
}\\
\subfloat[UMAP classification]{
\includegraphics[scale=0.52]{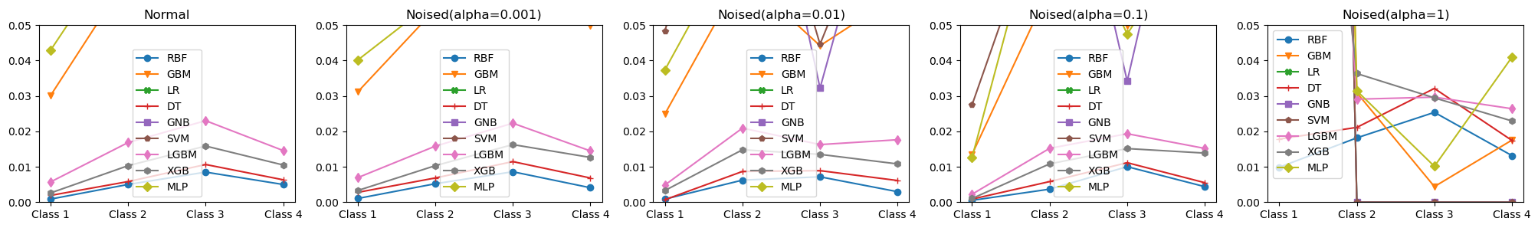}
\label{fig31}
}\\
\subfloat[t-SNE classification]{
\includegraphics[scale=0.52]{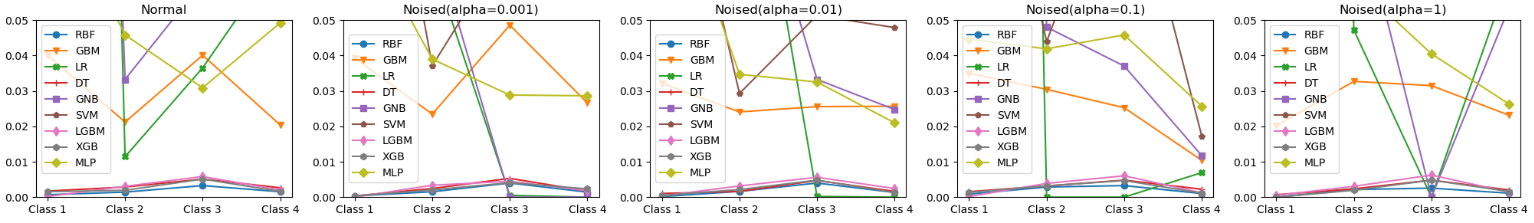}
\label{fig32}
}\\
\subfloat[Persistence Diagram classification]{
\includegraphics[scale=0.52]{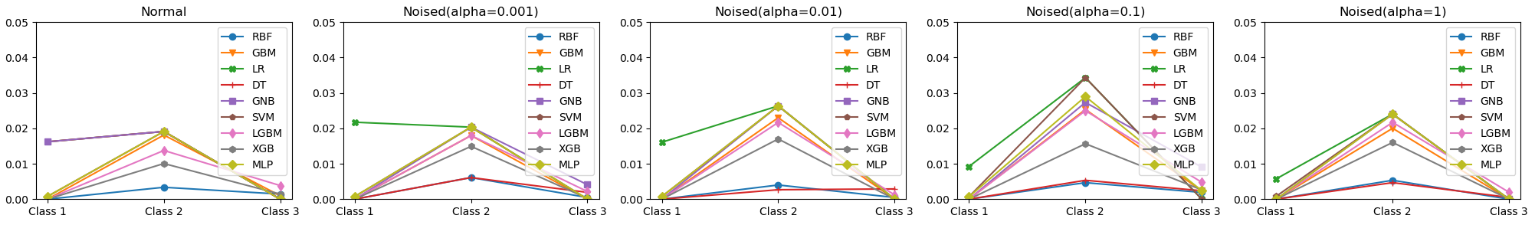}
\label{fig33}
}
\caption{False Positive Rate - CIC-MalMem-2022 dataset}
\label{fig3-cm}
\end{figure*}

\subsection{Machine Learning classification}\label{classif}

In order to answer a part of RQ2, the previous features extracted from existing techniques (PCA, t-SNE, UMAP) and TDA techniques (Mapper, Tomato, Persistence Diagram) are used for ML classification. We consider two cases: unsupervised learning and supervised learning. Before applying each case, we perform label encoding and min-max scaling (excepted classes) on the datasets.

\textbf{Unsupervised learning}. Features from Persistence Homology, Tomato, and TDA Mapper can be directly interpreted as classes. Each class can be represented by a cluster (graph, point). In the case of TDA Mapper, the most prominent graphs represent clusters. Tomato returns cluster points (most prominent peaks) by applying persistence diagrams on the node representation embeddings from graphs built on the datasets. Recall that using persistence diagrams, the homology groups track different information from the data representation embeddings in the datasets. The analysis of the number of malware samples appearing in a cluster, the number of malware samples associated to a malware class, and the relations between data features in the cluster and the original dataset can give some clues about the relation between a malware class and a given cluster.

\begin{figure*}[]
\centering
\subfloat[PCA classification]{
\includegraphics[scale=0.52]{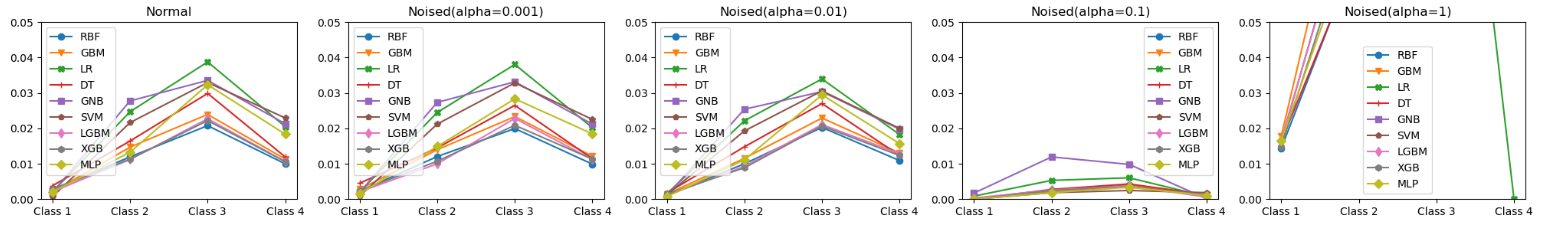}
\label{fig34}
}\\
\subfloat[UMAP classification]{
\includegraphics[scale=0.52]{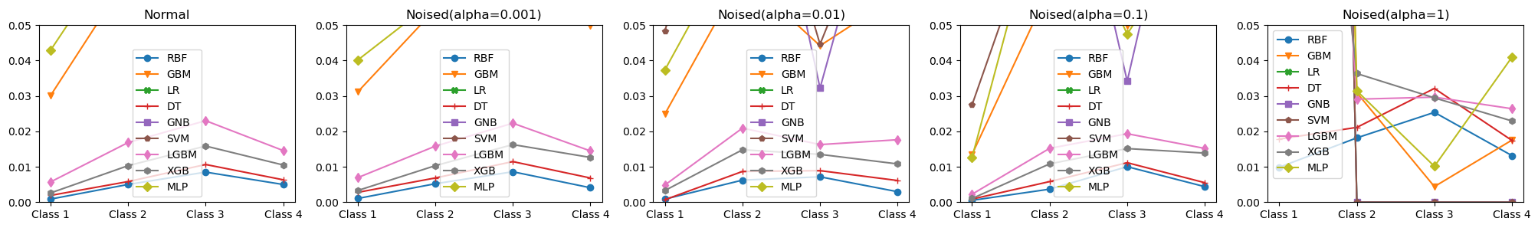}
\label{fig35}
}\\
\subfloat[t-SNE classification]{
\includegraphics[scale=0.52]{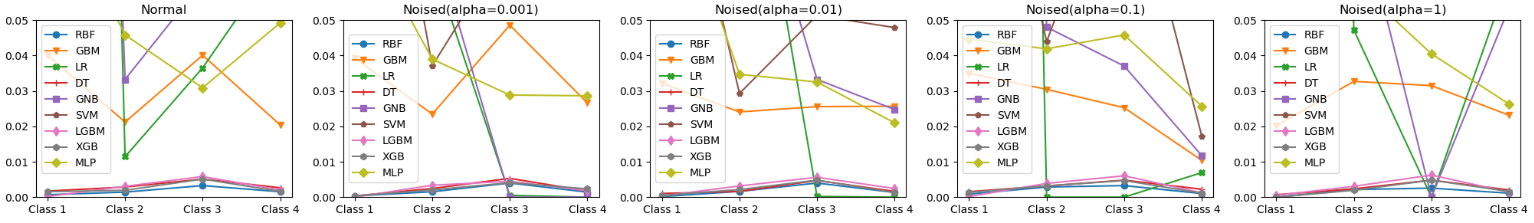}
\label{fig36}
}\\
\subfloat[Persistence Diagram classification]{
\includegraphics[scale=0.52]{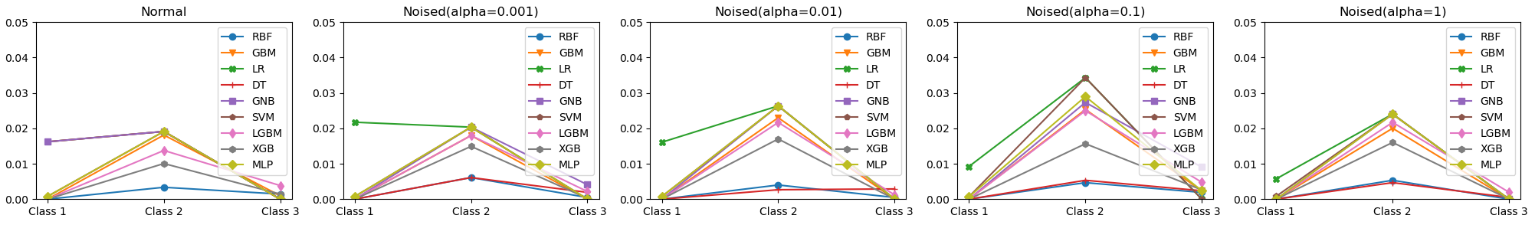}
\label{fig37}
}
\caption{False Positive Rate - CCCS-CIC-AndMal-2020 dataset}
\label{fig3-cca}
\end{figure*}

\textbf{Supervised learning}. PCA, t-SNE, UMAP, and Persistence Homology is used for supervised learning. For persistent diagrams, we used homology group features as classes (H0, H1, H2) and automatically relabeled data cloud points from these classes. TDA Mapper can be encoded to persistent diagrams or stochastic node embeddings~\cite{xu2021understanding} (e.g., DeepWalk, Node2Vec, or Struct2Vec) for classification. However, stochastic node embeddings return different outputs from different runs of the TDA Mapper algorithm on the same graph; thus making them non-reproducible. Then, we choose to encode TDA Mapper into persistence diagrams; since they are robust under pertubations, reproducible, and they are able to identify which predictors are more related to the outcome. Tomato already uses persistence diagrams to get exact prominent peaks representing the final predicted classes.

The extracted features are fed into different ML classifiers such as support vector machines (SVMs), logistic regression, random forest, decision tree, gradient boosting, xgboost, and lightgbm. 
These classifiers were already used in the literature for malware detection but the point is that we want to show that TDA techniques such as persistent diagrams can be used for classification like existing methods (PCA, t-SNE, UMAP), while achieving the same level of detection rate, less false positives, and even better. The next part of the RQ2 is to compare their execution time (i.e., feature extraction time, training time, inference time) and memory usage to see how they can be effectively apply in production settings.

\textbf{Hyper-parameter tuning}. We used the Optuna Library to search the best hyper parameters (e.g., learning rate, n\_estimators). For t-SNE, we vary to perplexity between 10 and 50 to select the best hyper parameter. We have selected the perplexity 20. For UMAP, we vary the number of neighbors between 3 and 7 and we have selected the value 5. t-SNE and UMAP were used as inputs of the TDA Mapper algorithm with the same hyper parameters. For ML classifiers, the learning rate is searched between 1e-03; and 1e-01. 

\begin{figure*}[]
\centering
\subfloat[PCA classification]{
\includegraphics[scale=0.52]{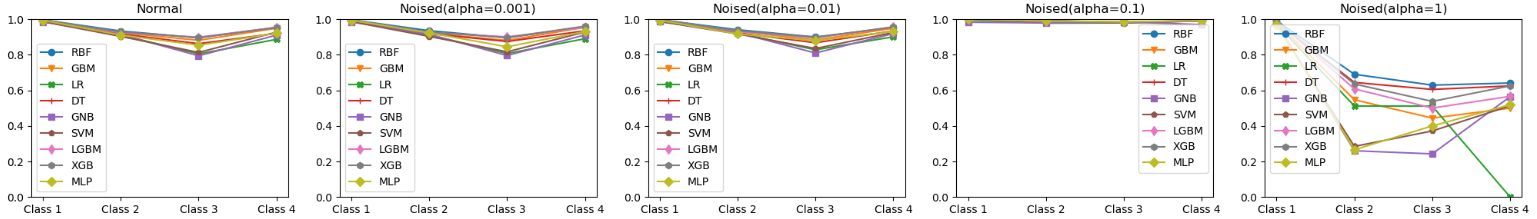}
\label{fig38}
}\\
\subfloat[UMAP classification]{
\includegraphics[scale=0.52]{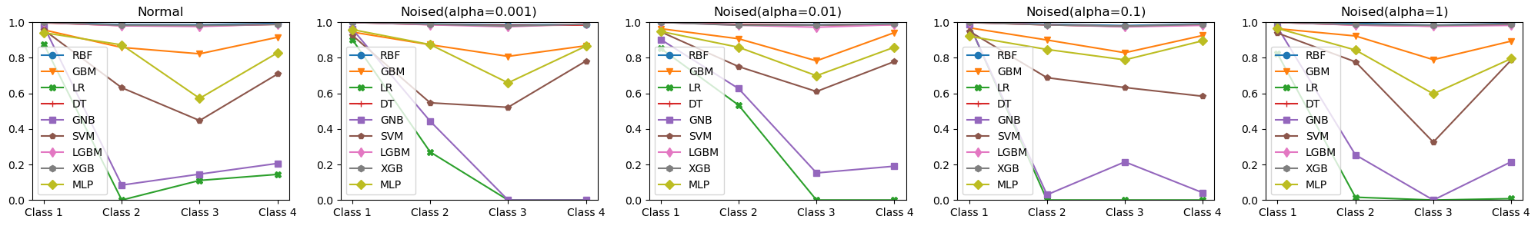}
\label{fig39}
}\\
\subfloat[t-SNE classification]{
\includegraphics[scale=0.52]{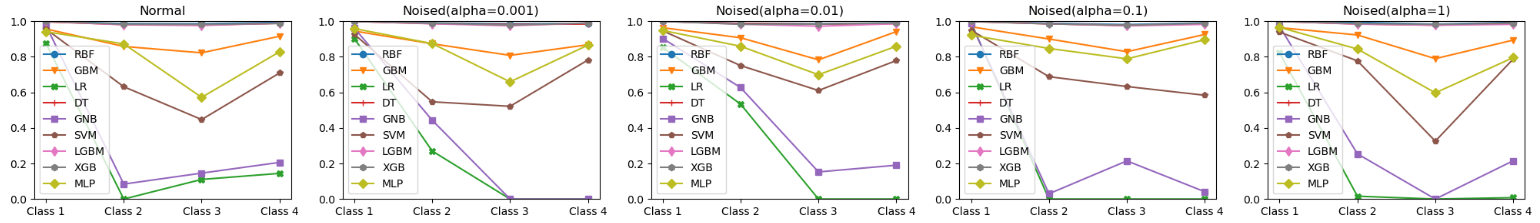}
\label{fig40}
}\\
\subfloat[Persistence Diagram classification]{
\includegraphics[scale=0.52]{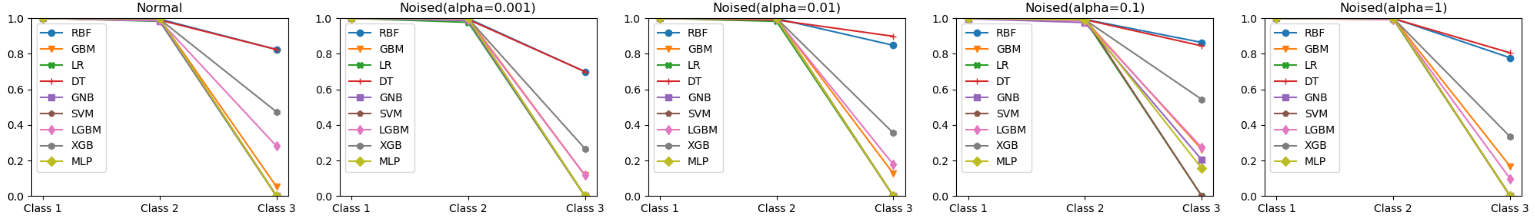}
\label{fig41}
}
\caption{Detection Rate - CIC-MalMem-2022 dataset}
\label{fig4-cm}
\end{figure*}

\begin{figure*}[]
\centering
\subfloat[PCA classification]{
\includegraphics[scale=0.5]{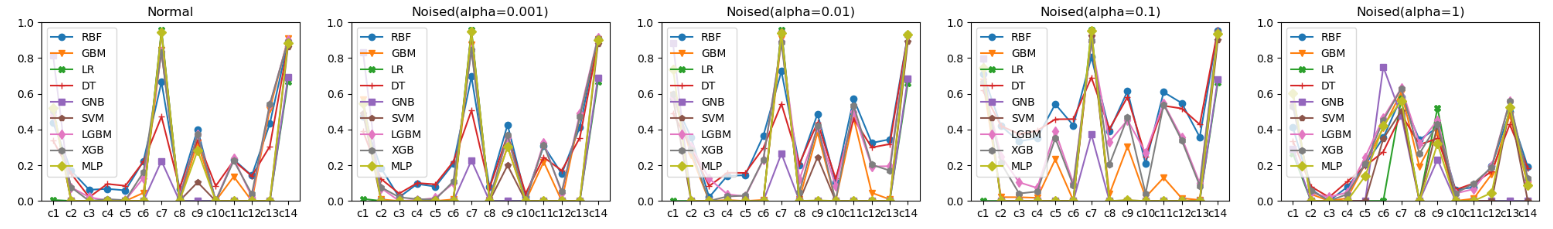}
\label{fig42}
}\\
\subfloat[UMAP classification]{
\includegraphics[scale=0.52]{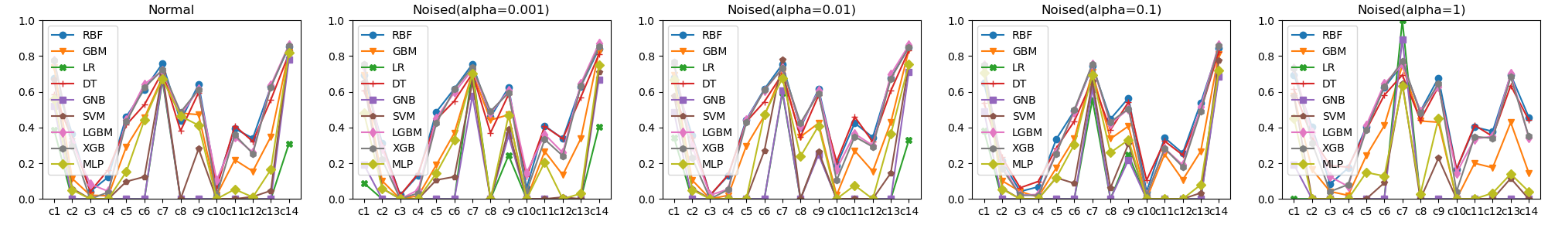}
\label{fig43}
}\\
\subfloat[t-SNE classification]{
\includegraphics[scale=0.52]{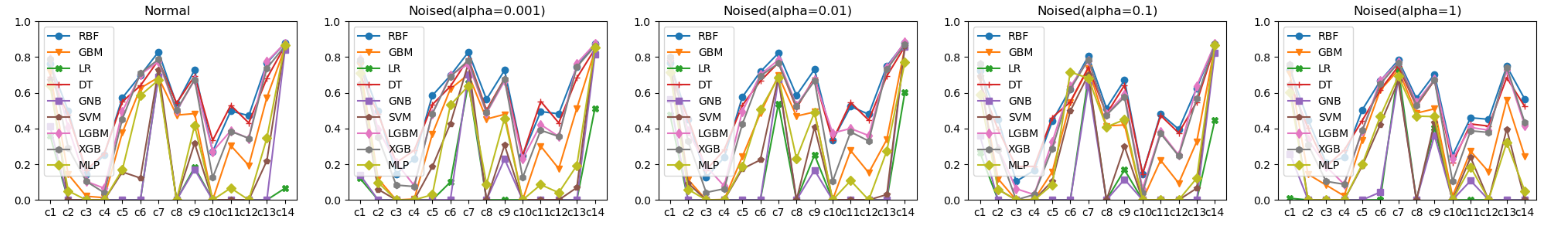}
\label{fig44}
}\\
\subfloat[Persistence Diagram classification]{
\includegraphics[scale=0.52]{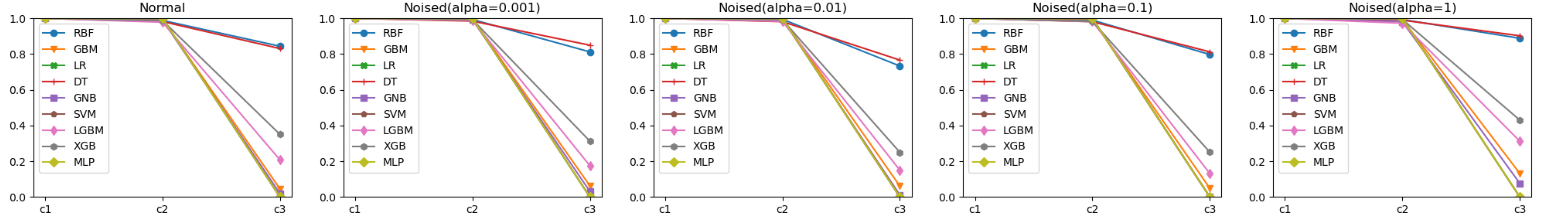}
\label{fig45}
}
\caption{Detection Rate - CCCS-CIC-AndMal-2020 dataset}
\label{fig4-cca}
\end{figure*}

\subsection{Evaluation Metrics}

\subsubsection{Malware Detection}

We use well-known metrics~\cite{8735821} such as detection rate and false positive rate to evaluate the performance of the studied techniques for malware detection. Other metrics such as precision and accuracy were not mentioned in the paper since the focus was on malware detection but they were computed during experiments and can be found in the replication package~\cite{tech-reportt}. 
Detection Rate (DR) is the probability that the ML model predicts correctly when the input sample is malicious (i.e., any malware category). False Positive Rate (FPR) is the probability that the ML model predicts the input sample as malicious (i.e., any malware category) although the input sample is benign. 

\begin{align*}
FPR&=\dfrac{FP}{FP+TN}  &  DR&=\dfrac{TP}{TP+FN}\\
\end{align*}
where False Positive (FP) is the number of benign samples misclassified as malicious (i.e., any class of malware), True Positive (TP) is the number of malicious samples correctly classified as malicious, False Negative (FN) is the number of malicious samples misclassified as benign, and True Negative (TN) is the number of benign samples correctly classified as benign.

In order to deal with RQ2, we compare the FPR of several classifiers (i.e., Random Forest, Gradient Boosting, Logistic Regression, Decision Tree, Gaussian NB, SVM, LightGBM, XGBoost, Multi-layer Perceptron) on features resulting from different methods (i.e., PCA, UMAP, TSNE, Persistence Diagram) on the CIC-MalMem-2022 and CCCS-CIC-AndMal-2020 datasets. This comparison is shown in Fig.~\ref{fig3-cm} and Fig.~\ref{fig3-cca}. For each FPR graph, we set the y-axis range between 0 and 5\%. In Fig.~\ref{fig4-cm} and Fig.~\ref{fig4-cca}, we also present the detection rate comparison between the previous classifiers on features resulting from different methods (PCA, UMAP, TSNE, Persistence Diagram) on the CIC-MalMem-2022 and CCCS-CIC-AndMal-2020 datasets. For each DR graph, we set the y-axis range between 0 and 1.

\subsubsection{Robustness}

In order to answer RQ3, the studied techniques are executed on datasets without noise (normal) and with the noise factor ranging between 0.001 to 0.1 (see Fig.~\ref{fig3-cm}-\ref{fig4-cca}). They are considered to be robust if the performance obtained on the datasets without noise (e.g., detection rate, false positive rate) remain approximately the same on the noised datasets.

\subsubsection{Execution}

For each dataset, we compute the memory usage, the feature extraction time, training time, and inference time using IPython time profilers ($\%time$, $\%timeit$) and memory profilers (e.g., $\%memit$). 

In order to answer RQ2, we record the CPU time, wall time, and memory usage for each feature extraction method on the CIC-MalMem-2022 and CCCS-CIC-AndMal-2020 datasets (see Table~\ref{fig-fe-1} and Table~\ref{fig-fe-2}). In Table~\ref{fig-fe-3}-\ref{fig-fe-10}, we also record the training and inference time/memory usage of the studied classifiers on features resulting from different methods (i.e., PCA, UMAP, TSNE, Persistence Diagram) on the CIC-MalMem-2022 and CCCS-CIC-AndMal-2020 datasets.

\section{Results}\label{evaluation}

In this section, we address the aforementioned research questions: (\textbf{RQ1}) the best feature engineering methods for malware analysis based on the criteria in~\cite{van1993categories}, execution time, and memory usage; (\textbf{RQ2}) the most performant techniques for malware detection in terms of detection rate, lowest false positive rate, lowest execution time, and lowest memory usage; and (\textbf{RQ3}) the most robust techniques against noise. 

\subsection{The best feature engineering techniques for malware analysis} 
We compare existing feature engineering techniques (i.e., PCA, t-SNE, UMAP) with TDA techniques such as TDA Mapper, Persistence Diagram, and Tomato based on the density of connected points (clusters), their uniqueness, their topology, the small within-cluster and large between-cluster dissimilarity, execution time, and memory usage. The results are shown below per evaluation criteria.

\subsubsection{High density of connected points} In Fig.~\ref{fig1}-\ref{fig2}, PCA returns 4 clusters representing feature classes (i.e., Trojan Horse, Benign, Ransomware, Spyware) with high density with no noise and the noise factor ranging from 0.001 to 0.1. When the noise factor is 1, the clusters are unstable which is identified by high perturbed data points. UMAP also outputs 4 clusters (Trojan Horse, Benign, Ransomware, Spyware) with normal data and noised data when the noise factor is between 0.001 and 0.1 (see Fig.~\ref{fig5}-\ref{fig6}). Unlike PCA and UMAP, t-SNE generates redundant and stochastic clusters thus making them hard to identify (see Fig.~\ref{fig3}-\ref{fig4}). t-SNE and UMAP clusters are aggregated while those with PCA are colinear or parallel. Unlike PCA/t-SNE/UMAP, TDA Mapper produces connected graphs of small/local clusters forming large/global clusters (see Fig.~\ref{fig7}-\ref{fig9} and Fig.~\ref{fig10}-\ref{fig12}). Cyber analysts can click on local clusters and see data details such as the number of data points (proportional to the sample size of the malware category) and the distribution of data in the cluster. Like PCA/t-SNE/UMAP, TDA techniques such as PCA TDA Mapper, UMAP TDA Mapper, and t-SNE TDA Mapper identified 4 clusters (Trojan Horse, Benign, Ransomware, Spyware) with normal data, and noised data when the noise factor is between 0.001 and 0.1. However, TDA Mapper is also perturbed when the noise factor is changed to 1 and there is no cluster formation. Unlike t-SNE/UMAP/PCA/TDA Mapper, persistence diagrams generates 3 features (homology groups) where the first cluster in blue are the number of connected components (0-betti numbers), the second cluster (in green) and the third cluster (in orange) are the holes in the point cloud data. When the noise factor is 1, the diagram show no points or a chaotic distribution (see Fig~\ref{fig15}-\ref{fig13}). Like PCA/UMAP/t-SNE, Tomato identified 4 $\delta$-prominent peaks (clusters) with normal data and under noise when factor is 0.001 on the CIC-MalMem-2022 dataset (see Fig.~\ref{fig18}). However, Tomato is perturbed from 0.01 to 1; it can be observed by the disproportional number of clusters. On the CCCS-CIC-AndMal-2020 dataset, Tomato identified 3 topological features with normal data and under noise when $\alpha$ is between 0.001 to 0.1 (see Fig.~\ref{fig17}).   

\subsubsection{Uniqueness of clusters} For PCA and UMAP, we have 4 unique clusters on the CIC-MalMem-2022 and CCCS-CIC-AndMal-2020 datasets with normal and noised data ranging between 0.001 and 0.1 (see Fig.~\ref{fig1}-\ref{fig2} and Fig.~\ref{fig5}-\ref{fig6}). For t-SNE, we have no unique clusters (see Fig.~\ref{fig3}-\ref{fig4}). Similarly to PCA and UMAP, TDA Mapper produces 4 unique clusters that are reproducible with normal and noised data when the noise factor is between 0.001 and 0.1 (see Fig.~\ref{fig7}-\ref{fig9} and Fig.~\ref{fig10}-\ref{fig12}). Persistence diagrams return 3 unique topological features with normal and noised data (see Fig~\ref{fig15}-\ref{fig13}). Like PCA and UMAP, Tomato outputs unique clusters with normal and noised data ranging between 0.001 and 0.1 (see Fig.~\ref{fig18}-\ref{fig17}). The uniqueness of clusters is particularly important to identify particular features related to some malware families (e.g., port call on 3389 for RDP ransomware attacks).

\subsubsection{Cluster topology} 
On the CIC-MalMem-2022 and CCCS-CIC-AndMal-2020 datasets, PCA has a more linear topology compared to UMAP and t-SNE (see Fig.~\ref{fig1}-\ref{fig2}). UMAP topology is close to ellipsoids while t-SNE has no specific topology since its distribution is stochastic (see Fig.~\ref{fig5}-\ref{fig6} and Fig.~\ref{fig3}-\ref{fig4}). Unlike PCA/UMAP/t-SNE, clusters in PCA/UMAP/t-SNE TDA Mapper have a nerve topology that can be either linear, ellipsoid, star, or geoid (see Fig.~\ref{fig7}-\ref{fig9} and Fig.~\ref{fig10}-\ref{fig12}). 

\subsubsection{Small within-cluster dissimilarity}
PCA, UMAP, and t-SNE have a smaller within-cluster distance than TDA Mapper on the CIC-MalMem-2022 and CCCS-CIC-AndMal-2020 datasets. This is explained by the fact that TDA Mapper creates edges between cluster vertices (see Fig.~\ref{fig7}-\ref{fig9} and Fig.~\ref{fig10}-\ref{fig12}). Like PCA/UMAP/t-SNE, persistence diagrams have a smaller within-cluster dissimilarity on the CIC-MalMem-2022 and CCCS-CIC-AndMal-2020 datasets. In practice, malware analysts can assign new malware samples (unknown samples) to an existing cluster (e.g., Ransomware) based on their small within-cluster distance; thus allowing to identify the class/category of the new malwares. 
\subsubsection{Large between-cluster dissimilarity}
On the CIC-MalMem-2022 and CCCS-CIC-AndMal-2020 datasets, PCA has a larger between-cluster distance than t-SNE and UMAP. t-SNE and UMAP have a more compact structure, then their between-cluster distance is zero. Like PCA, TDA Mapper and Tomato have a significant between-cluster dissimilarity on the CIC-MalMem-2022 and CCCS-CIC-AndMal-2020 datasets.  A large between-cluster dissimilarity enables malware analysts to better classify new malware samples.  
\begin{table}[h]
\centering
\caption{Feature Extraction Performance - CIC-MalMem-2022 dataset}
\label{fig-fe-1}
\resizebox{\columnwidth}{!}{
\begin{tabular}{|l|l|l|l|}
\hline
Unsupervised                & \textbf{CPU time (s)} & \textbf{Wall time (s)} & \textbf{memory (MiB)} \\ \hline
\textbf{PCA}                & 1.08                  & 0.27                   & 58.25                       \\ \hline
\textbf{UMAP}               & 150                   & 23.9                   & 341.16                      \\ \hline
\textbf{TSNE}               & 2483                  & 253                    & 228.55                      \\ \hline
\textbf{Pers. Diagram} & 24.7                  & 21.4                   & 2273.05                     \\ \hline
\textbf{Tomato}             & 37.8                  & 37.8                   & 140.20                      \\ \hline
\textbf{TDAMapper-PCA}        & 30.4                    & 30.4                   & 135.77                    \\ \hline
\textbf{TDAMapper-UMAP}       & 112                   & 16.7                   & 228.83                      \\ \hline
\textbf{TDAMapper-TSNE}       & 2555                  & 255                    & 140.20    \\ \hline                
\end{tabular}
}
\end{table}

\begin{table}[h]
\caption{Feature Extraction Performance - CCCS-CIC-AndMal-2020}
\label{fig-fe-2}
\centering
\resizebox{\columnwidth}{!}{
\begin{tabular}{|l|l|l|l|}
\hline
Unsupervised                & \textbf{CPU time (s)} & \textbf{Wall time (s)} & \textbf{memory (MiB)} \\ \hline
\textbf{PCA}                & 2.89                  & 0.48                   & 63.73                       \\ \hline
\textbf{UMAP}               & 83                   & 9.76                   & 441.23                      \\ \hline
\textbf{TSNE}               & 2981                 & 284                    &  234.21                    \\ \hline
\textbf{Pers. Diagram} & 26.3                  & 18.3                   & 1283.90                     \\ \hline
\textbf{Tomato}             & 77                  & 77                   & 198.93                      \\ \hline
\textbf{TDAMapper-PCA}        & 19                    & 2.65                   & 55.45                   \\ \hline
\textbf{TDAMapper-UMAP}       & 101                   & 12.4                   & 155.28                    \\ \hline
\textbf{TDAMapper-TSNE}       & 2473                  & 238                    & 135.77    \\ \hline                
\end{tabular}
}
\end{table}

\subsubsection{Execution time and space}

In terms of feature extraction time, PCA is the fastest, followed by Persistence Diagram on the CIC-MalMem-2022 dataset (see Table~\ref{fig-fe-1}). On the CCCS-CIC-AndMal-2020 dataset, PCA is still the fastest, followed by TDAMapper-PCA (see Table~\ref{fig-fe-2}). In terms of feature extraction space, PCA has the lowest memory usage, followed by Tomato and TDAMapper-TSNE (i.e., TDA Mapper algorithm with t-SNE for feature extraction) on the CIC-MalMem-2022 dataset. On the CCCS-CIC-AndMal-2020 dataset, TDAMapper-PCA has the lowest memory usage, followed by PCA. Overall, PCA and TDA Mapper (using PCA and TSNE) are the best feature engineering techniques. 

\begin{boxblock}{Summary 1}
    \begin{itemize}
      \item For clustering, Malware analysts can use PCA and TDA Mapper (with PCA for feature extraction); since they have a high density of connected points, cluster uniqueness, small within-cluster and large between-cluster dissimilarity, relatively low execution time and low memory usage: \textit{1) CCCS-CIC-AndMal-2020:} 2.89s and 63.73 MiB for PCA, 19s and 55.45 MiB for TDAMapper-PCA; \textit{2) CIC-MalMem-2022:} 1.08s and 58.25 MiB for PCA, 30.4s and 135.77 MiB for TDAMapper-PCA.
      
      \item Malware analysts can also use Persistence Diagram, UMAP, and t-SNE to precisely identify overlapping malware features/clusters while considering the processing time: \textit{1) CIC-MalMem-2022:} 24.7s for Persistence Diagram, 150s for UMAP, and 2483s for t-SNE; \textit{2) CCCS-CIC-AndMal-2020:} 26.3s for Persistence Diagram, 83s for UMAP, and 2981s for t-SNE.
      
      \item Malware analysts can use TDA Mapper (with PCA/UMAP/t-SNE for feature extraction) to accurately identify links between malware features/clusters while considering the processing time and space: \textit{1) CCCS-CIC-AndMal-2020:} 19s and 55.45 MiB for TDAMapper-PCA, 101s and 155.28 MiB for TDAMapper-UMAP, 2473s and 135.77 MiB for TDAMapper-TSNE; \textit{2) CIC-MalMem-2022:} 30.4s and 135.77 MiB for TDAMapper-PCA, 112s and 228.83 MiB for TDAMapper-UMAP, 2555s and 140.20 MiB for TDAMapper-TSNE.
    \end{itemize}
\end{boxblock}

\subsection{The most performant techniques for malware detection} 
We compare existing feature engineering techniques 
(i.e., PCA, t-SNE, UMAP, Persistence Diagram) using different classifiers (i.e., Random Forest, Gradient Boosting, Logistic Regression, Decision Tree, Gaussian NB, SVM, LightGBM, XGBoost, Multi-layer Perceptron) based on performance metrics such as false positive rate, detection rate, execution time, and memory usage. The results are shown below per performance metric.

\subsubsection{False positive rate}

Overall, Random Forest and Decision Tree achieved the best performance with a false positive rate (FPR) under 0.6\% on average using Persistence Diagram and t-SNE for feature extraction on the CIC-MalMem-2022 and CCCS-CIC-AndMal-2020 datasets when processing normal data and noised data for any value of $\alpha$ (see Fig.~\ref{fig3-cm} and Fig.~\ref{fig3-cca}). Using all the feature engineering techniques (i.e., PCA, UMAP, t-SNE, Persistence Diagram), XGBoost and LightGBM achieved a good performance with a FPR under 2.5\% on average on both datasets when processing normal data and noised data for $\alpha$ between 0.001 and 0.1 (see Fig.~\ref{fig3-cm} and Fig.~\ref{fig3-cca}). When the noised data is proportional to normal data (i.e., $\alpha$ is equals to 1), XGBoost and LightGBM FPRs remain under 2.5\% on average using Persistence Diagram and t-SNE on all datasets but the FPRs quickly increase up to 5\% using PCA and UMAP. With other classifiers such as Linear Regression and SVM, the false positive rate is very unstable and increase by up to 5\% using UMAP and t-SNE with normal and noised data (for any value of $\alpha$). In practice, malware detectors deal with huge amount of traffic and generate several alerts that are at least 80\% not malicious and very few are malicious. Thus, Persistence Diagram and t-SNE can help malware detectors to generate less FPR and help cyber analysts to considerably reduce the time of the malware analysis.

\begin{table}[h]
\caption{PCA Classification Performance- CIC-MalMem-2022 dataset}
\label{fig-fe-3}
\centering
\resizebox{\columnwidth}{!}{
\begin{tabular}{|l|lll|lll|}
\hline
\multirow{2}{*}{\begin{tabular}[c]{@{}l@{}}Supervised (PCA\\ Feature Extraction)\\ \end{tabular}} & \multicolumn{3}{c|}{\textbf{Training}}                                                        & \multicolumn{3}{l|}{\textbf{Inference}}                                                      \\ \cline{2-7} 
                                                                                                            & \multicolumn{1}{l|}{\textbf{CPU (s)}} & \multicolumn{1}{l|}{\textbf{Wall (s)}} & \textbf{Mem(MiB)} & \multicolumn{1}{l|}{\textbf{CPU(s)}} & \multicolumn{1}{l|}{\textbf{Wall(s)}} & \textbf{Mem (MiB)} \\ \hline
\textbf{RandomForest}                                                                                       & \multicolumn{1}{l|}{3.84}             & \multicolumn{1}{l|}{3.84}              & 31.18             & \multicolumn{1}{l|}{0.125}           & \multicolumn{1}{l|}{0.134}            & 0.00               \\ \hline
\textbf{XGBoost}                                                                                            & \multicolumn{1}{l|}{103}              & \multicolumn{1}{l|}{8.94}              & 5.44              & \multicolumn{1}{l|}{0.188}           & \multicolumn{1}{l|}{0.022}            & 0.00               \\ \hline
\textbf{LightGBM}                                                                                           & \multicolumn{1}{l|}{6.47}             & \multicolumn{1}{l|}{0.597}             & 0.00              & \multicolumn{1}{l|}{0.484}           & \multicolumn{1}{l|}{0.044}            & 0.05               \\ \hline
\textbf{Decision Tree}                                                                                      & \multicolumn{1}{l|}{0.109}            & \multicolumn{1}{l|}{0.112}             & 0.04              & \multicolumn{1}{l|}{0}               & \multicolumn{1}{l|}{0.003}            & 0.00               \\ \hline
\textbf{Logistic Reg.}                                                                                      & \multicolumn{1}{l|}{0.312}            & \multicolumn{1}{l|}{0.291}             & 2.14              & \multicolumn{1}{l|}{0}               & \multicolumn{1}{l|}{0.001}            & 0.00               \\ \hline
\textbf{MLP}                                                                                                & \multicolumn{1}{l|}{42}               & \multicolumn{1}{l|}{42}                & 0.02              & \multicolumn{1}{l|}{0.016}           & \multicolumn{1}{l|}{0.012}            & 8.95               \\ \hline
\textbf{GBM}                                                                                                & \multicolumn{1}{l|}{16.5}             & \multicolumn{1}{l|}{16.5}              & 6.21              & \multicolumn{1}{l|}{0.047}           & \multicolumn{1}{l|}{0.064}            & 0.00               \\ \hline
\textbf{Gaussian NB}                                                                                        & \multicolumn{1}{l|}{0.031}            & \multicolumn{1}{l|}{0.012}             & 0.00              & \multicolumn{1}{l|}{0}               & \multicolumn{1}{l|}{0.002}            & 0.00               \\ \hline
\textbf{SVM}                                                                                                & \multicolumn{1}{l|}{8.97}             & \multicolumn{1}{l|}{8.98}              & 199.38            & \multicolumn{1}{l|}{9.05}            & \multicolumn{1}{l|}{9.05}             & 0.29               \\ \hline
\end{tabular}
}
\end{table}

\begin{table}[h]
    \caption{UMAP Classification Performance- CIC-MalMem-2022 dataset}
    \label{fig-fe-4}
    \centering
    \resizebox{\columnwidth}{!}{
    \begin{tabular}{|l|lll|lll|}
    \hline
    \multirow{2}{*}{\begin{tabular}[c]{@{}l@{}}Supervised (UMAP\\ Feature Extraction)\\\end{tabular}} & \multicolumn{3}{c|}{\textbf{Training}}                                                        & \multicolumn{3}{l|}{\textbf{Inference}}                                                      \\ \cline{2-7} 
                                                                                                                 & \multicolumn{1}{l|}{\textbf{CPU (s)}} & \multicolumn{1}{l|}{\textbf{Wall (s)}} & \textbf{Mem(MiB)} & \multicolumn{1}{l|}{\textbf{CPU(s)}} & \multicolumn{1}{l|}{\textbf{Wall(s)}} & \textbf{Mem (MiB)} \\ \hline
    \textbf{RandomForest}                                                                                        & \multicolumn{1}{l|}{4.05}             & \multicolumn{1}{l|}{4.04}              & 27.31             & \multicolumn{1}{l|}{0.156}           & \multicolumn{1}{l|}{0.148}            & 0.02               \\ \hline
    \textbf{XGBoost}                                                                                             & \multicolumn{1}{l|}{127}              & \multicolumn{1}{l|}{11.2}              & 3.04              & \multicolumn{1}{l|}{0.344}           & \multicolumn{1}{l|}{0.029}            & 0.00               \\ \hline
    \textbf{LightGBM}                                                                                            & \multicolumn{1}{l|}{6.16}             & \multicolumn{1}{l|}{0.541}             & 8.95              & \multicolumn{1}{l|}{0.844}           & \multicolumn{1}{l|}{0.067}            & 0.10               \\ \hline
    \textbf{Decision Tree}                                                                                       & \multicolumn{1}{l|}{0.094}            & \multicolumn{1}{l|}{0.1}               & 0.87              & \multicolumn{1}{l|}{0}               & \multicolumn{1}{l|}{0.001}            & 0.00               \\ \hline
    \textbf{Logistic Reg.}                                                                                       & \multicolumn{1}{l|}{0.266}            & \multicolumn{1}{l|}{0.253}             & 2.29              & \multicolumn{1}{l|}{0}               & \multicolumn{1}{l|}{0.001}            & 0.00               \\ \hline
    \textbf{MLP}                                                                                                 & \multicolumn{1}{l|}{27.1}             & \multicolumn{1}{l|}{27.1}              & 0.88              & \multicolumn{1}{l|}{0}               & \multicolumn{1}{l|}{0.010}            & 8.95               \\ \hline
    \textbf{GBM}                                                                                                 & \multicolumn{1}{l|}{15.8}             & \multicolumn{1}{l|}{15.9}              & 2.42              & \multicolumn{1}{l|}{0.063}           & \multicolumn{1}{l|}{0.061}            & 0.62               \\ \hline
    \textbf{Gaussian NB}                                                                                         & \multicolumn{1}{l|}{0}                & \multicolumn{1}{l|}{0.009}             & 2.42              & \multicolumn{1}{l|}{0.016}           & \multicolumn{1}{l|}{0.005}            & 0.62               \\ \hline
    \textbf{SVM}                                                                                                 & \multicolumn{1}{l|}{42.6}             & \multicolumn{1}{l|}{42.6}              & 206.9             & \multicolumn{1}{l|}{25.3}            & \multicolumn{1}{l|}{25.3}             & 0.29               \\ \hline
    \end{tabular}
    }
    \end{table}

\subsubsection{Detection rate}

Overall, Random Forest, Decision Tree, XGBoost, and LightGBM achieved the highest detection rate between 99\% and 100\% with UMAP and t-SNE on the CIC-MalMem-2022 dataset when processing normal data and noised data for any value of $\alpha$ (see Fig.~\ref{fig4-cm}). However, their detection rates become unstable (i.e., ranging between 0 to 90\%) when the noised data is proportional to normal data (i.e., $\alpha$ is equal to 1). Random Forest and Decision Tree achieved a high detection rate over 80\% on average with Persistence Diagram for feature extraction on the CIC-MalMem-2022 dataset when processing normal and noised data (for any value of $\alpha$). Using UMAP and t-SNE, other classifiers such as Linear Regression and Gaussian NB have a variable detection rate that can decrease under 1\% with normal and noised data. 

On the CCCS-CIC-AndMal-2020 dataset, Random Forest and Decision Tree achieved the best performance with feature extraction methods such as PCA, UMAP, t-SNE, and Persistence Diagram. However, the average detection rate is under 40\% because the dataset is imbalanced; which makes detection performance variable between classes (see Fig.~\ref{fig4-cca}). Nonetheless, we observe that the performance of Random Forest and Decision Tree using persistence diagrams on the imbalanced dataset is affected just like that of the others. In Section~\ref{discussions}, we will discuss the results and show how it can be improved.

        \begin{table}[h]
            \caption{Persistence Diagram Classification Performance- CIC-MalMem-2022 dataset}
    \label{fig-fe-6}
    \centering
    \resizebox{\columnwidth}{!}{
    \begin{tabular}{|l|lll|lll|}
    \hline
    \multirow{2}{*}{\begin{tabular}[c]{@{}l@{}}Supervised (with \\ Persistence Diagram)\\ \end{tabular}} & \multicolumn{3}{c|}{\textbf{Training}}                                                        & \multicolumn{3}{l|}{\textbf{Inference}}                                                      \\ \cline{2-7} 
                                                                                                     & \multicolumn{1}{l|}{\textbf{CPU (s)}} & \multicolumn{1}{l|}{\textbf{Wall (s)}} & \textbf{Mem(MiB)} & \multicolumn{1}{l|}{\textbf{CPU(s)}} & \multicolumn{1}{l|}{\textbf{Wall(s)}} & \textbf{Mem (MiB)} \\ \hline
    \textbf{RandomForest}                                                                            & \multicolumn{1}{l|}{0.641}            & \multicolumn{1}{l|}{0.643}             & 0.07              & \multicolumn{1}{l|}{0.031}           & \multicolumn{1}{l|}{0.038}            & 0.00               \\ \hline
    \textbf{XGBoost}                                                                                 & \multicolumn{1}{l|}{16.3}             & \multicolumn{1}{l|}{1.43}              & 0.18              & \multicolumn{1}{l|}{0}               & \multicolumn{1}{l|}{0.008}            & 0.00               \\ \hline
    \textbf{LightGBM}                                                                                & \multicolumn{1}{l|}{3.17}             & \multicolumn{1}{l|}{0.304}             & 3.09              & \multicolumn{1}{l|}{0.172}           & \multicolumn{1}{l|}{0.02}             & 0.00               \\ \hline
    \textbf{Decision Tree}                                                                           & \multicolumn{1}{l|}{0.016}            & \multicolumn{1}{l|}{0.012}             & 0.01              & \multicolumn{1}{l|}{0}               & \multicolumn{1}{l|}{0.002}            & 0.00               \\ \hline
    \textbf{Logistic Reg.}                                                                           & \multicolumn{1}{l|}{0.063}            & \multicolumn{1}{l|}{0.069}             & 0.00              & \multicolumn{1}{l|}{0}               & \multicolumn{1}{l|}{0.001}            & 0.00               \\ \hline
    \textbf{MLP}                                                                                     & \multicolumn{1}{l|}{9.08}             & \multicolumn{1}{l|}{9.09}              & 0.00              & \multicolumn{1}{l|}{0.016}           & \multicolumn{1}{l|}{0.006}            & 3.21               \\ \hline
    \textbf{GBM}                                                                                     & \multicolumn{1}{l|}{2.56}             & \multicolumn{1}{l|}{2.56}              & 1.33              & \multicolumn{1}{l|}{0.016}           & \multicolumn{1}{l|}{0.014}            & 0.00               \\ \hline
    \textbf{Gaussian NB}                                                                             & \multicolumn{1}{l|}{0}                & \multicolumn{1}{l|}{0.003}             & 0.23              & \multicolumn{1}{l|}{0}               & \multicolumn{1}{l|}{0.001}            & -0.23              \\ \hline
    \textbf{SVM}                                                                                     & \multicolumn{1}{l|}{0.562}            & \multicolumn{1}{l|}{0.584}             & 18.21             & \multicolumn{1}{l|}{0.469}           & \multicolumn{1}{l|}{0.475}            & 0.20               \\ \hline
    \end{tabular}
    }
    \end{table}
    
    \begin{table}[h]
        \caption{Persistence Diagram Classification Performance- CCCS-CIC-AndMal-2020}
    \label{fig-fe-10}
    \centering
    \resizebox{\columnwidth}{!}{
    \begin{tabular}{|l|lll|lll|}
    \hline
    \multirow{2}{*}{\begin{tabular}[c]{@{}l@{}}Supervised (with\\ Persistence Diagram)\\ \end{tabular}} & \multicolumn{3}{c|}{\textbf{Training time}}                                                        & \multicolumn{3}{l|}{\textbf{Inference time}}                                                      \\ \cline{2-7} 
                                                                                                                & \multicolumn{1}{l|}{\textbf{CPU (s)}} & \multicolumn{1}{l|}{\textbf{Wall (s)}} & \textbf{Mem(MiB)} & \multicolumn{1}{l|}{\textbf{CPU(s)}} & \multicolumn{1}{l|}{\textbf{Wall(s)}} & \textbf{Mem (MiB)} \\ \hline
    \textbf{RandomForest}                                                                                       & \multicolumn{1}{l|}{0.734}             & \multicolumn{1}{l|}{0.746}              & 37.31             & \multicolumn{1}{l|}{0.047}           & \multicolumn{1}{l|}{0.039}            & 0.02               \\ \hline
    \textbf{XGBoost}                                                                                            & \multicolumn{1}{l|}{18}              & \multicolumn{1}{l|}{1.56}              & 0.19             & \multicolumn{1}{l|}{0}           & \multicolumn{1}{l|}{0.017}            & 0.03               \\ \hline
    \textbf{LightGBM}                                                                                           & \multicolumn{1}{l|}{3.7}             & \multicolumn{1}{l|}{0.359}             & 0.05             & \multicolumn{1}{l|}{0.172}           & \multicolumn{1}{l|}{0.024}            & 0               \\ \hline
    \textbf{Decision Tree}                                                                                      & \multicolumn{1}{l|}{0.016}            & \multicolumn{1}{l|}{0.016}             & 0             & \multicolumn{1}{l|}{0}               & \multicolumn{1}{l|}{0.002}            & 0               \\ \hline
    \textbf{Logistic Reg.}                                                                                      & \multicolumn{1}{l|}{0.063}            & \multicolumn{1}{l|}{0.081}             & 0           & \multicolumn{1}{l|}{0}               & \multicolumn{1}{l|}{0.002}            & 0               \\ \hline
    \textbf{MLP}                                                                                                & \multicolumn{1}{l|}{7.48}               & \multicolumn{1}{l|}{7.5}                & 0              & \multicolumn{1}{l|}{0}           & \multicolumn{1}{l|}{0.006}            & 3.36              \\ \hline
    \textbf{GBM}                                                                                                & \multicolumn{1}{l|}{2.84}             & \multicolumn{1}{l|}{2.84}              & 0.20              & \multicolumn{1}{l|}{0.016}           & \multicolumn{1}{l|}{0.019}            & 0               \\ \hline
    \textbf{Gaussian NB}                                                                                        & \multicolumn{1}{l|}{0}            & \multicolumn{1}{l|}{0.005}             & 0.00              & \multicolumn{1}{l|}{0}               & \multicolumn{1}{l|}{0.003}            & 0              \\ \hline
    \textbf{SVM}                                                                                                & \multicolumn{1}{l|}{0.750}             & \multicolumn{1}{l|}{0.741}              & 25.12            & \multicolumn{1}{l|}{0.688}            & \multicolumn{1}{l|}{0.698}             & 0               \\ \hline
    \end{tabular}
    }
    \end{table}
    
\begin{table}[h]
    \caption{TSNE Classification Performance- CIC-MalMem-2022 dataset}
    \label{fig-fe-5}
    \centering
    \resizebox{\columnwidth}{!}{
    \begin{tabular}{|l|lll|lll|}
    \hline
    \multirow{2}{*}{\begin{tabular}[c]{@{}l@{}}Supervised (TSNE\\ Feature Extraction)\\\end{tabular}} & \multicolumn{3}{c|}{\textbf{Training}}                                                        & \multicolumn{3}{l|}{\textbf{Inference}}                                                      \\ \cline{2-7} 
                                                                                                                 & \multicolumn{1}{l|}{\textbf{CPU (s)}} & \multicolumn{1}{l|}{\textbf{Wall (s)}} & \textbf{Mem(MiB)} & \multicolumn{1}{l|}{\textbf{CPU(s)}} & \multicolumn{1}{l|}{\textbf{Wall(s)}} & \textbf{Mem(MiB)} \\ \hline
    \textbf{RandomForest}                                                                                        & \multicolumn{1}{l|}{3.7}              & \multicolumn{1}{l|}{3.71}              & 0.00              & \multicolumn{1}{l|}{0.141}           & \multicolumn{1}{l|}{0.145}            & 0.21               \\ \hline
    \textbf{XGBoost}                                                                                             & \multicolumn{1}{l|}{125}              & \multicolumn{1}{l|}{11.1}              & 14.31             & \multicolumn{1}{l|}{0.281}           & \multicolumn{1}{l|}{0.031}            & 0.09               \\ \hline
    \textbf{LightGBM}                                                                                            & \multicolumn{1}{l|}{6.7}              & \multicolumn{1}{l|}{0.6}               & 5.40              & \multicolumn{1}{l|}{0.86}            & \multicolumn{1}{l|}{0.075}            & 0.00               \\ \hline
    \textbf{Decision Tree}                                                                                       & \multicolumn{1}{l|}{0.094}            & \multicolumn{1}{l|}{0.1}               & 0.00              & \multicolumn{1}{l|}{0}               & \multicolumn{1}{l|}{0.002}            & 0.00               \\ \hline
    \textbf{Logistic Reg.}                                                                                       & \multicolumn{1}{l|}{0.188}            & \multicolumn{1}{l|}{0.181}             & 0.12              & \multicolumn{1}{l|}{0}               & \multicolumn{1}{l|}{0.001}            & 0.00               \\ \hline
    \textbf{MLP}                                                                                                 & \multicolumn{1}{l|}{25.6}             & \multicolumn{1}{l|}{25.6}              & 0.33              & \multicolumn{1}{l|}{0}               & \multicolumn{1}{l|}{0.010}            & 5.78               \\ \hline
    \textbf{GBM}                                                                                                 & \multicolumn{1}{l|}{15.6}             & \multicolumn{1}{l|}{15.6}              & 27.28             & \multicolumn{1}{l|}{0.063}           & \multicolumn{1}{l|}{0.065}            & 0.00               \\ \hline
    \textbf{Gaussian NB}                                                                                         & \multicolumn{1}{l|}{0}                & \multicolumn{1}{l|}{0.009}             & 0.00              & \multicolumn{1}{l|}{0}               & \multicolumn{1}{l|}{0.003}            & 0.00               \\ \hline
    \textbf{SVM}                                                                                                 & \multicolumn{1}{l|}{42.2}             & \multicolumn{1}{l|}{42.3}              & 174.36            & \multicolumn{1}{l|}{24.9}            & \multicolumn{1}{l|}{24.9}             & 0.04               \\ \hline
    \end{tabular}
    }
    \end{table}

\subsubsection{Execution time and space}

In terms of training time, Decision Tree and Gaussian NB are the fastest followed by Logistic Regression and Random Forest on the CIC-MalMem-2022 and CCCS-CIC-AndMal-2020 datasets using PCA, UMAP, t-SNE, and Persistence Diagram for feature extraction (see Table~\ref{fig-fe-3}-\ref{fig-fe-6} and Table~\ref{fig-fe-7}-\ref{fig-fe-10}). In terms of training space, Gaussian NB and LightGBM have the lowest memory usage followed by Multi-Layer Perceptron (MLP) and Decision Tree on the CIC-MalMem-2022 dataset using PCA for feature extraction (see Table~\ref{fig-fe-3}). Using UMAP for feature extraction, Decision Tree and MLP have the lowest memory usage followed by Gradient Boosting Machine (GBM) and Gaussian NB on the CIC-MalMem-2022 dataset. Using t-SNE, Decision Tree, Random Forest, and Gaussian NB have the lowest memory usage on the CIC-MalMem-2022 dataset. Using Persistence Diagram, Logistic Regression, Multi-Layer Perceptron, Decision Tree, Gaussian NB, and Random Forest have the lowest memory usage on the CIC-MalMem-2022 dataset. On the CCCS-CIC-AndMal-2020 dataset, Gaussian NB, MLP, Decision Tree and Logistic Regression have the lowest memory usage using PCA, UMAP, t-SNE, and Persistence Diagram for feature extraction (see Table~\ref{fig-fe-7}-\ref{fig-fe-10}).  
    
    \begin{table}[h]
    \centering
    \caption{PCA Classification Performance- CCCS-CIC-AndMal-2020}
    \label{fig-fe-7}
    \resizebox{\columnwidth}{!}{
    \begin{tabular}{|l|lll|lll|}
    \hline
    \multirow{2}{*}{\begin{tabular}[c]{@{}l@{}}Supervised (PCA\\ Feature Extraction)\\ \end{tabular}} & \multicolumn{3}{c|}{\textbf{Training}}                                                        & \multicolumn{3}{l|}{\textbf{Inference}}                                                      \\ \cline{2-7} 
                                                                                                                & \multicolumn{1}{l|}{\textbf{CPU (s)}} & \multicolumn{1}{l|}{\textbf{Wall (s)}} & \textbf{Mem(MiB)} & \multicolumn{1}{l|}{\textbf{CPU(s)}} & \multicolumn{1}{l|}{\textbf{Wall(s)}} & \textbf{Mem (MiB)} \\ \hline
    \textbf{RandomForest}                                                                                       & \multicolumn{1}{l|}{7.97}             & \multicolumn{1}{l|}{7.98}              & 501.57             & \multicolumn{1}{l|}{0.344}           & \multicolumn{1}{l|}{0.338}            & 2.29               \\ \hline
    \textbf{XGBoost}                                                                                            & \multicolumn{1}{l|}{443}              & \multicolumn{1}{l|}{38.1}              & 33.20             & \multicolumn{1}{l|}{0.891}           & \multicolumn{1}{l|}{0.09}            & 0.00               \\ \hline
    \textbf{LightGBM}                                                                                           & \multicolumn{1}{l|}{19.8}             & \multicolumn{1}{l|}{1.74}             & 13.86              & \multicolumn{1}{l|}{1.55}           & \multicolumn{1}{l|}{0.146}            & 1.14               \\ \hline
    \textbf{Decision Tree}                                                                                      & \multicolumn{1}{l|}{0.203}            & \multicolumn{1}{l|}{0.201}             & 9.36             & \multicolumn{1}{l|}{0}               & \multicolumn{1}{l|}{0.004}            & 1.14               \\ \hline
    \textbf{Logistic Reg.}                                                                                      & \multicolumn{1}{l|}{2.86}            & \multicolumn{1}{l|}{1.29}             & 9.13            & \multicolumn{1}{l|}{0}               & \multicolumn{1}{l|}{0.002}            & 2.29               \\ \hline
    \textbf{MLP}                                                                                                & \multicolumn{1}{l|}{243}               & \multicolumn{1}{l|}{80}                & 0.00              & \multicolumn{1}{l|}{0.016}           & \multicolumn{1}{l|}{0.015}            & 9.30              \\ \hline
    \textbf{GBM}                                                                                                & \multicolumn{1}{l|}{56.9}             & \multicolumn{1}{l|}{57}              & 18.25              & \multicolumn{1}{l|}{0.156}           & \multicolumn{1}{l|}{0.156}            & 1.14               \\ \hline
    \textbf{Gaussian NB}                                                                                        & \multicolumn{1}{l|}{0.016}            & \multicolumn{1}{l|}{0.012}             & 0.00              & \multicolumn{1}{l|}{0}               & \multicolumn{1}{l|}{0.008}            & 2.29              \\ \hline
    \textbf{SVM}                                                                                                & \multicolumn{1}{l|}{95}             & \multicolumn{1}{l|}{95}              & 226.34            & \multicolumn{1}{l|}{41.8}            & \multicolumn{1}{l|}{41.9}             & 0.00               \\ \hline
    \end{tabular}
    }
    \end{table}
    
    \begin{table}[h]
        \caption{UMAP Classification Performance- CCCS-CIC-AndMal-2020}
    \label{fig-fe-8}
    \centering
    \resizebox{\columnwidth}{!}{
    \begin{tabular}{|l|lll|lll|}
    \hline
    \multirow{2}{*}{\begin{tabular}[c]{@{}l@{}}Supervised (UMAP\\ Feature Extraction)\\ \end{tabular}} & \multicolumn{3}{c|}{\textbf{Training}}                                                        & \multicolumn{3}{l|}{\textbf{Inference}}                                                      \\ \cline{2-7} 
                                                                                                                & \multicolumn{1}{l|}{\textbf{CPU (s)}} & \multicolumn{1}{l|}{\textbf{Wall (s)}} & \textbf{Mem(MiB)} & \multicolumn{1}{l|}{\textbf{CPU(s)}} & \multicolumn{1}{l|}{\textbf{Wall(s)}} & \textbf{Mem (MiB)} \\ \hline
    \textbf{RandomForest}                                                                                       & \multicolumn{1}{l|}{7.39}             & \multicolumn{1}{l|}{7.39}              & 370.47             & \multicolumn{1}{l|}{0.328}           & \multicolumn{1}{l|}{0.335}            & 1.34               \\ \hline
    \textbf{XGBoost}                                                                                            & \multicolumn{1}{l|}{426}              & \multicolumn{1}{l|}{37.3}              & 33.14             & \multicolumn{1}{l|}{0.516}           & \multicolumn{1}{l|}{0.54}            & 0.00               \\ \hline
    \textbf{LightGBM}                                                                                           & \multicolumn{1}{l|}{21.1}             & \multicolumn{1}{l|}{1.81}             & 18.51             & \multicolumn{1}{l|}{1.59}           & \multicolumn{1}{l|}{0.161}            & 1.14               \\ \hline
    \textbf{Decision Tree}                                                                                      & \multicolumn{1}{l|}{21.1}            & \multicolumn{1}{l|}{1.81}             & 5.16             & \multicolumn{1}{l|}{1.59}               & \multicolumn{1}{l|}{0.161}            & 1.14               \\ \hline
    \textbf{Logistic Reg.}                                                                                      & \multicolumn{1}{l|}{1.89}            & \multicolumn{1}{l|}{1.32}             & 9.14           & \multicolumn{1}{l|}{0}               & \multicolumn{1}{l|}{0.002}            & 2.29               \\ \hline
    \textbf{MLP}                                                                                                & \multicolumn{1}{l|}{176}               & \multicolumn{1}{l|}{68}                & 0.00              & \multicolumn{1}{l|}{0.016}           & \multicolumn{1}{l|}{0.011}            & 4.08              \\ \hline
    \textbf{GBM}                                                                                                & \multicolumn{1}{l|}{54.7}             & \multicolumn{1}{l|}{54.7}              & 18.25              & \multicolumn{1}{l|}{0.141}           & \multicolumn{1}{l|}{0.146}            & 1.14               \\ \hline
    \textbf{Gaussian NB}                                                                                        & \multicolumn{1}{l|}{0.00}            & \multicolumn{1}{l|}{0.01}             & 0.00              & \multicolumn{1}{l|}{0}               & \multicolumn{1}{l|}{0.008}            & 2.29              \\ \hline
    \textbf{SVM}                                                                                                & \multicolumn{1}{l|}{114}             & \multicolumn{1}{l|}{114}              & 55.77            & \multicolumn{1}{l|}{38.9}            & \multicolumn{1}{l|}{39}             & 0.00               \\ \hline
    \end{tabular}
    }
    \end{table}

\begin{table}[h]
        \caption{t-SNE Classification Performance- CCCS-CIC-AndMal-2020}
    \label{fig-fe-9}
    \centering
    \resizebox{\columnwidth}{!}{
    \begin{tabular}{|l|lll|lll|}
    \hline
    \multirow{2}{*}{\begin{tabular}[c]{@{}l@{}}Supervised (TSNE\\ Feature Extraction)\\ \end{tabular}} & \multicolumn{3}{c|}{\textbf{Training}}                                                        & \multicolumn{3}{l|}{\textbf{Inference}}                                                      \\ \cline{2-7} 
                                                                                                                & \multicolumn{1}{l|}{\textbf{CPU (s)}} & \multicolumn{1}{l|}{\textbf{Wall (s)}} & \textbf{Mem(MiB)} & \multicolumn{1}{l|}{\textbf{CPU(s)}} & \multicolumn{1}{l|}{\textbf{Wall(s)}} & \textbf{Mem (MiB)} \\ \hline
    \textbf{RandomForest}                                                                                       & \multicolumn{1}{l|}{6.19}             & \multicolumn{1}{l|}{6.19}              & 318.63             & \multicolumn{1}{l|}{0.297}           & \multicolumn{1}{l|}{0.299}            & 1.14               \\ \hline
    \textbf{XGBoost}                                                                                            & \multicolumn{1}{l|}{408}              & \multicolumn{1}{l|}{38.3}              & 34.96             & \multicolumn{1}{l|}{0.953}           & \multicolumn{1}{l|}{0.096}            & 0.00               \\ \hline
    \textbf{LightGBM}                                                                                           & \multicolumn{1}{l|}{22.1}             & \multicolumn{1}{l|}{1.99}             & 10.14             & \multicolumn{1}{l|}{2.09}           & \multicolumn{1}{l|}{0.202}            & 1.14               \\ \hline
    \textbf{Decision Tree}                                                                                      & \multicolumn{1}{l|}{0.172}            & \multicolumn{1}{l|}{0.167}             & 4.83             & \multicolumn{1}{l|}{0}               & \multicolumn{1}{l|}{0.003}            & 1.14               \\ \hline
    \textbf{Logistic Reg.}                                                                                      & \multicolumn{1}{l|}{2.09}            & \multicolumn{1}{l|}{1.31}             & 9.13           & \multicolumn{1}{l|}{0}               & \multicolumn{1}{l|}{0.002}            & 2.27               \\ \hline
    \textbf{MLP}                                                                                                & \multicolumn{1}{l|}{169}               & \multicolumn{1}{l|}{70}                & 0.00              & \multicolumn{1}{l|}{0.016}           & \multicolumn{1}{l|}{0.01}            & 4.08              \\ \hline
    \textbf{GBM}                                                                                                & \multicolumn{1}{l|}{58.8}             & \multicolumn{1}{l|}{58.9}              & 40.35              & \multicolumn{1}{l|}{0.172}           & \multicolumn{1}{l|}{0.165}            & 1.14               \\ \hline
    \textbf{Gaussian NB}                                                                                        & \multicolumn{1}{l|}{0.016}            & \multicolumn{1}{l|}{0.011}             & 0.00              & \multicolumn{1}{l|}{0.016}               & \multicolumn{1}{l|}{0.009}            & 2.29              \\ \hline
    \textbf{SVM}                                                                                                & \multicolumn{1}{l|}{103}             & \multicolumn{1}{l|}{103}              & 69.28            & \multicolumn{1}{l|}{38.1}            & \multicolumn{1}{l|}{38.2}             & 0.00               \\ \hline
    \end{tabular}
    }
    \end{table}
    
In terms of inference time, Decision Tree, Gaussian NB, Logistic Regression, and MLP are the fastest on the CIC-MalMem-2022 dataset using PCA, UMAP, and t-SNE for feature extraction (see Table~\ref{fig-fe-3}-\ref{fig-fe-6}). Using persistence diagrams, XGBoost, Decision Tree, Gaussian NB, and Logistic Regression have the fastest inference time on the CIC-MalMem-2022 dataset (see Table~\ref{fig-fe-6}). On the CCCS-CIC-AndMal-2020 dataset, Decision Tree, Logistic Regression, Gaussian NB, and MLP are the fastest using PCA and t-SNE for feature extraction (see Table~\ref{fig-fe-7} and Table~\ref{fig-fe-9}). 

Using UMAP, Logistic Regression, Gaussian NB, and MLP have the fastest inference time compared to other classifiers such as SVM and LightGBM on the CCCS-CIC-AndMal-2020 dataset (see Table~\ref{fig-fe-8}). Using persistence diagrams, ML classifiers such as XGBoost, Decision Tree, Logistic Regression, MLP, and Gaussian NB have the fastest inference time on the CCCS-CIC-AndMal-2020 dataset (see Table~\ref{fig-fe-10}). 

In terms of inference space, XGBoost, SVM, Decision Tree, LightGBM, and Gradient Boosting Machine (GBM) have the lowest memory usage than other classifiers such as MLP and Random Forest on the CIC-MalMem-2022 dataset using PCA, UMAP, and t-SNE (see Table~\ref{fig-fe-3}-\ref{fig-fe-6}). Using persistence diagrams, all classifiers except MLP have a good memory usage on the CIC-MalMem-2022 dataset. On the CCCS-CIC-AndMal-2020 dataset, XGBoost and SVM have the lowest memory usage followed by Decision Tree, Random Forest, LightGBM, and GBM using PCA, UMAP, and t-SNE (see Table~\ref{fig-fe-7}-\ref{fig-fe-10}). Using persistence diagrams, all classifiers except MLP have a good memory usage on the CCCS-CIC-AndMal-2020 dataset.

\begin{boxblock}{Summary 2}
    \begin{itemize}
      \item Malware analysts can use Random Forest and Decision Tree with t-SNE and Persistence Diagram as feature extractors for malware detection even when there is too much noise in input data; since \textit{1) they have a false positive rate under 0.6\% on average} on the CIC-MalMem-2022 and CCCS-CIC-AndMal-2020 datasets with high pertubations; \textit{2) they have the best detection rate} on the CIC-MalMem-2022 dataset: 99-100\% with t-SNE and 80-100\% with Persistence Diagram; \textit{3) their training time and space is relatively low} on both datasets: 0.016-0.734s and 0-37.31 MiB with Persistence Diagram, 0.094-6.19s and 0-318.63 MiB with t-SNE; and \textit{4) their inference time and space is relatively low} on both datasets: 0.031-0.047s and 0-0.02 MiB with Persistence Diagram, 0-0.297s and 0-1.14 MiB with t-SNE.
      
      \item Malware analysts can also use XGBoost and LightGBM with feature engineering technique such as t-SNE and UMAP for malware detection when there is a medium noise in input data; since \textit{1) they have a false positive rate under 2.5\% on average} on both datasets with the noise factor between 0.001 and 0.1; \textit{2) they have the best detection rate} on the CIC-MalMem-2022 dataset (e.g., 99-100\% with t-SNE and UMAP); \textit{3) their training time and space is relatively medium} on both datasets: 6.16-426s and 3.04-33.14 MiB with UMAP, 6.7-408s and 5.4-34.96 MiB with t-SNE; and \textit{4) their inference time and space is relatively low} on both datasets: 0.344-1.59s and 0-1.14 MiB with UMAP, 0.281-2.09s and 0-1.14 MiB with t-SNE.
      
      \item The detection performance of Random Forest and Decision Tree using persistence diagrams on imbalanced datasets (e.g., CCCS-CIC-AndMal-2020) is affected just like that of the existing techniques (i.e., PCA, UMAP, and t-SNE).
      
    \end{itemize}
\end{boxblock}

\subsection{The most robust techniques against noisy input data}

In terms of supervised classification, Random Forest and Decision Tree are robust against noise for all cases (i.e., any value of $\alpha$) when Persistence Diagram and t-SNE are used for feature extraction on the CIC-MalMem-2022 and CCCS-CIC-AndMal-2020 datasets (see Fig.~\ref{fig3-cm}-\ref{fig3-cca} and Fig.~\ref{fig4-cm}-\ref{fig4-cca}). XGboost and LightGBM does not perform well with persistence diagrams but they are robust with t-SNE on both datasets. Using PCA and UMAP, Random Forest, Decision Tree, XGboost, and LightGBM are highly perturbed when $\alpha$ is equal to 1. Other classifiers such as Linear Regression, GBM, and MLP are robust using Persistence Diagram but they are highly perturbed using t-SNE, PCA, and UMAP. 

In terms of unsupervised classification, PCA, UMAP, and t-SNE are robust against noise on the CIC-MalMem-2022 and CCCS-CIC-AndMal-2020 datasets when $\alpha$ is between 0.001 and 0.1. When $\alpha$ is equal to 1, they are unable to achieve clustering. Like PCA/UMAP/t-SNE, TDA Mapper, Persistence Diagram, and Tomato are robust between when $\alpha$ is between 0.001 and 0.1 but they are also perturbed when the noised data is proportional to normal data (i.e., $\alpha$ is equal to 1). 

\begin{boxblock}{Summary 3}
    \begin{itemize}
      \item In supervised classification, Malware analysts can use Random Forest and Decision Tree with t-SNE and Persistence Diagram for robustness against very noised input data, which is an important factor to deal with malwares in real-world traffic or logs that are usually very noisy. XGBoost and LightGBM can only be used with t-SNE for robustness (see Section~\ref{discussions}).
      
      \item In unsupervised classification, Malware analysts can use TDA Mapper (with PCA for feature extraction), t-SNE, Persistence Diagram, and Tomato on input data with a medium noise since their robustness can be affected on very noised inputs. 
    \end{itemize}
\end{boxblock}

\section{Discussion}
\label{discussions}

For malware analysis, PCA is able to precisely extract malware features from data with low memory and processing time but it is not able to identify hidden relationships between those features~\cite{10.5555/3491440.3492052}. TDA techniques such as TDA Mapper generates a graph of small/local clusters forming large/global clusters allowing to easily identify hidden relations between malware features. For example, the ransomware cluster identified in the normal case (see Fig.~\ref{fig7}) forms a graph of local connected clusters, showing relations between sets of malware samples (local clusters) based on their topological features. Malware analysts can click on local clusters and see data details such as the number of data points and the distribution of data in the cluster. 
TDA Mapper is fully customizable with its dynamic multi-resolution glimpse feature allowing deeper analysis of the structure and information behind the generation process of the malware cluster graphs~\cite{10.5555/3491440.3492052}. 

Like UMAP and t-SNE, persistence diagrams are able to identify overlapping malware features which is an important factor to analyze common PEs in malware binaries (e.g., byte sequences, opcodes, network calls). In real-world, it is often common to have some malwares that belong to several classes because they have multiple behaviors. For example, they can use some ransomware, trojan-dropper, backdoor, and spyware techniques at the same time. Thus, persistent diagrams and t-SNE can help malware analysts to recognize these overlapping malware behaviors. 
Unlike UMAP and t-SNE, persistence diagrams use topological features (the number of connected components, the holes in the point cloud data) to identify overlapping malware features. In addition, persistence diagrams, Tomato, and TDA Mapper (with PCA for feature extraction) have low execution time compared to UMAP and t-SNE on the studied datasets. Persistence diagrams consume more memory than the other techniques but it can be improved to run at scale~\cite{lacombe2018large}. 

For malware detection, PCA coupled with classifiers such as Random Forest, Decision Tree, XGBoost, and LightGBM generates significant false positives with an FPR up to 2\% for normal inputs and over 5\% with very noise data (i.e., $\alpha$ is equal to 1). Using t-SNE and persistence diagrams for malware detection, Random Forest, Decision Tree, XGBoost, and LightGBM have a FPR under 0.6\% with normal and noised inputs. However, other classifiers such as Gaussian NB and Logistic Regression produces too many false positives when they are coupled with UMAP and t-SNE, while they are more robust with persistence diagrams. This is particularly important for malware detectors (e.g., endpoint detection responses, antivirus) to accurately identify malicious patterns in real-world traffic or logs that are usually very noisy. 

Detection performance is good (i.e., between 99\% and 100\%) when t-SNE and UMAP are coupled with classifiers such as Random Forest, Decision Tree, XGBoost, and LightGBM on the CIC-MalMem-2022 dataset. However, the training and inference processing time of these classifiers are significant with t-SNE and UMAP compared to Persistence Diagram and PCA. With persistence diagrams, Random Forest and Decision Tree have a good detection rate between 80\% to 100\% but the detection rate quickly decreases with other classifiers such as XGBoost and LightGBM on the CIC-MalMem-2022 dataset. Thus, malware analysts must only use persistence diagrams with Random Forest and Decision Tree for malware detection. 

On imbalanced datasets like CCCS-CIC-AndMal-2020, the detection performance is good for some malware classes and worse for others with the studied classifiers using PCA, UMAP, t-SNE, and Persistence Diagram. For example, the fourteenth class in Fig.~\ref{fig44} is predicted with an accuracy between 80 and 85\% by some classifiers while the accuracy for the third class is very low (i.e., between 0\% and 20\%). 
Nonetheless, we observe that Decision Tree and Random Forest still achieved the best performance with Persistence Diagram, t-SNE, and UMAP. Since real-world malware detectors deal with very imbalanced datasets, we wanted to observe how TDA techniques behave on those datasets compared to existing techniques (i.e., PCA, UMAP, t-SNE). From the analysis, the detection performance of the studied classifiers using TDA techniques like persistence diagrams are affected on imbalanced datasets just like that of those using PCA, UMAP, and t-SNE. In practice, the traditional way to improve detection on imbalanced datasets is SMOTE~\cite{chawla2002smote}; it helps resampling the datasets by adding more malware examples (oversampling) or randomly deleting them for majority classes. 

\section{Recommendations}
\label{recommendations}

In this section, we propose some recommendations about deploying TDA models for network/log detection and malware analysis in single-server, multi-server, small-network, large-network, and multi-region environments. 

In a single-server environment, we have different logs generated by applications, services, and the OS. Host-based detectors can be implemented using random forest (RF) and decision tree (DT) with persistent diagrams (PD) and t-SNE to efficiently analyze noised logs (e.g., Winevt, syslog) and detect malware activities (e.g., registry modification, file deletion) with low FPR, high detection rate, and low overhead. Malware analysts can use multi-threading mechanisms to take advantage of the computer resources to train models on remote servers since host-based detectors must continuously learn from data to improve their performance. 

In a multi-server environment, training servers in the cloud must be auto scaled to deal with the huge number of events received from multiple servers to re-train models. When the workload becomes high, autoscaling will help to dynamically scale up/down the number of training instances provisioned for the proposed ML models (i.e., RF with PD/TSNE, DT with PD/TSNE) and an application load balancing can be used to distribute log streams across training instances. Distributed versions of Random Forest and decision trees~\cite{biau2016random, chen2016parallel} coupled with distributed persistent diagrams~\cite{bauer2014distributed,lacombe2018large} can also help to efficiently optimize the detection. More often, output alerts from host detectors can be huge even when they are accurate. Alerts can be efficiently correlated using a TDAMapper-PCA service (i.e., TDA Mapper with PCA as feature extractor) that will generate a graph clusters of alerts, allowing to pinpoint high level alerts. The service based on TDA Mapper can be also auto scaled to deal with tremendous alerts and an another application load balancer can be also used upfront to manage alert workloads. 

In a small network, we have inbound/outbound traffic exchanged through the network. Network-based detectors can be also implemented using Random Forest (RF) and Decision Tree (DT) with persistent diagrams (PD) and t-SNE to efficiently analyze packet sessions from a network tap, mirror ports, or a bastion server with low FPR, high detection rate, and low overhead. In the cloud, VPC flow logs can be streamlined to the network detection service based on the proposed models, to analyze IP flows and identify anomalous/infected IP addresses exploited by malwares. TDAMapper-PCA can be also used to generate graph clusters of IP addresses~\cite{gutierrez2018cyber} to simplify cyber analyst tasks and to help them identify potentially infected subnets controlled/targeted by malwares. Once an infected subnet is detected, cyber analysts can automatically take proper actions (e.g., using serverless functions) to shutdown/stop machines in the subnets. 

In a large network, network load balancers will be used to distribute traffic workloads across several network nodes. At each node, network detectors based on the optimized versions of the proposed models~\cite{biau2016random, lacombe2018large, chen2016parallel, bauer2014distributed} can be deployed as a service, on network taps, or bastion servers to analyze the inbound/outbound traffic and generate alerts when an anomaly is detected. To deal with huge alerts from multiple network nodes, network/log alerts from a local node can be first aggregated using TDAMapper-PCA to identify high level alert clusters, that will be further sent to a central node for clustering using TDAMapper-PCA. The processing services running TDAMapper-PCA must be auto scaled to ensure high performance and application load balancers can be also used upfront to handle big alerts. 

In a multi-region environment, network and log events come from multiple countries across the globe. Geographically distributed systems such as content deliver networks (CDNs) will be used to faster deliver network/log events by caching their contents in proxy servers close to network/host detectors than initial servers. This will allow an efficient analysis and detection of malicious activities across multiple regions. Malware analysts can also use network and application load balancers after the CDN to respectively manage network traffic and log workloads. However, we will have tremendous alerts generated from multiple detectors. Application load balancers can be applied to distribute alerts across multiple correlation services based on TDAMapper-PCA. Correlation services running TDAMapper-PCA must be auto scaled as well to ensure high availability and performance.

\section{Threats to validity}~\label{threat2valid}

This section discusses threats to validity that can affect our study. 


\textbf{Internal validity} threats concern any confounding factor that could influence our results.  Some Kernel methods~\cite{reininghaus2015stable,kusano2017kernel} have been proposed  to make TDA features consumable by machine learning classifiers but they are not yet practically applicable, limiting to scope of experimentation in this paper. During experimentation, some techniques like t-SNE were time-consuming to execute and reproduce. It took 1 month to run ML classifiers for each feature engineering technique and to repeat 2 to 3 times the execution to ensure the correctness of the results.

\textbf{External validity} threats concern the possibility of generalizing our results. TDA techniques have been extensively used in different domains (e.g., infectious diseases~\cite{chen2021topological, bleher2021topology}, renewable energy~\cite{bush2021topological}). In this study, we evaluated the performance of TDA techniques (with/without ML classifiers) for malware detection using two datasets from different platforms (i.e., PC, mobile) and with different properties (i.e., balanced, imbalanced). For large-scale systems, we have proposed some recommendations to scale the best-identified techniques for malware detection on different environments (e.g., multi-server, network, multi-region) but further validation is necessary to put these recommendations in practice.   

\textbf{Reliability validity} threats concern the possibility of replicating this study. We have provided all the necessary details needed to replicate the study in~\cite{tech-reportt}.

\textbf{Construct validity} threats concern the relationship between the theory and the observation. In our study, well-known metrics such as false positive rate and detection rate have been used for performance evaluation between the studied techniques. Other metrics such as precision and accuracy were not mentioned in the paper since the focus was on malware detection and analysis but they were computed during experiments and can be found in the replication package~\cite{tech-reportt}.

\section{Conclusion}~\label{conclusion}

In this work, we used TDA techniques (TDA Mapper, Tomato, Persistence Diagram) for malware detection and compared them against the existing feature engineering techniques (PCA, t-SNE, UMAP) on different ML classifiers. Some recommendations have been also proposed to efficiently deploy the best-identified models in corporate environments. Results show that TDA Mapper (with PCA as feature extractor) is better for clustering and for identifying hidden relationships between malware clusters compared to PCA. Persistent diagrams are better to identify overlapping malware clusters with low execution time compared to UMAP and t-SNE. For malware detection, malware analysts can use Random Forest and Decision Tree with t-SNE and Persistent Diagram to achieve better performance and robustness on noisy data. XGBoost and LightGBM can only be used with t-SNE to achieve better performance and robustness. In the future, we plan to execute TDA techniques on streaming data (network traffic, logs) for malware detection.

\section*{Acknowledgment}

This work is partly funded by the Fonds de Recherche du Québec (FRQ), Natural Sciences and Engineering Research Council of Canada (NSERC), Canadian Institute for Advanced Research (CIFAR), and Mathematics of Information Technology and Complex Systems (MITACS).



\bibliographystyle{IEEEtran}
\balance
\bibliography{main}

\end{document}